\documentclass[fleqn,usenatbib]{mnras}

\usepackage{newtxtext,newtxmath}

\usepackage[T1]{fontenc}

\DeclareRobustCommand{\VAN}[3]{#2}
\let\VANthebibliography\thebibliography
\def\thebibliography{\DeclareRobustCommand{\VAN}[3]{##3}\VANthebibliography}

\usepackage{graphicx}	
\usepackage{amsmath}	

\title[Group velocity of oblique Alfv{\'e}n waves]{Group velocity of obliquely propagating Alfv{\'e}n waves in a magnetized dusty plasma}

\author[L. B. De Toni, R. Gaelzer, L. F. Ziebell]{
L. B. De Toni,$^{1}$\thanks{E-mail: luan.toni@ufrgs.br}
R. Gaelzer$^{1}$\thanks{E-mail: rudi.gaelzer@ufrgs.br}
and L. F. Ziebell$^{1}$\thanks{E-mail: luiz.ziebell@ufrgs.br}
\\
$^{1}$Instituto de F{\'i}sica, Universidade Federal do Rio Grande do Sul, CP 15051, 91501-970, Porto Alegre, RS, Brazil
}

\date{This is a pre-copyedited, author-produced PDF of an article accepted for publication in Monthly Notices of the Royal Astronomical Society following peer review.}

\pubyear{2022}

\begin{document}
\label{firstpage}
\pagerange{\pageref{firstpage}--\pageref{lastpage}}
\maketitle

\begin{abstract}
In this work we investigate the characteristics of the group velocity of obliquely propagating Alfv{\'e}n waves in a dusty plasma typical of a stellar wind. The dispersion relation is derived with the aid of the kinetic theory for a magnetized dusty plasma consisting of electrons and ions, with distribution of momenta described by a Maxwellian function. The dust particles are considered to be immobile and have all the same size; they are electrically charged by absorption of plasma particles via inelastic collisions and by photoionization. We numerically solve the dispersion relation and calculate the components of group velocity (along and transverse to the magnetic field) for the normal modes, namely the compressional and shear Alfv{\'e}n waves (CAW and SAW). The results show that the direction of the group velocity of CAWs is greatly modified with the wave-vector direction. On the other hand, SAWs will present group velocity propagating practically along the magnetic field. The changes in dust parameters, such as number density and equilibrium electrical charge, may significantly change the waves' characteristics. It is seen that for sufficiently high dust to ion number density ratio, the SAWs may present perpendicular group velocity propagating in opposite direction to the perpendicular phase velocity, in a small interval of wavenumber values; we also notice that this interval may change, or even vanish, when the flux of radiation incident on the dust is altered, changing the equilibrium electrical charge of the grains.
\end{abstract}




\section{Introduction}
Space plasmas are commonly populated by dust particles with very high masses relative to an ion, and a wide range of sizes, with radii of the order of nanometres to a few millimetres. These grains of dust will accumulate their own electrical charge, which can be positive or negative, depending on the charging mechanisms and surrounding environment. The presence of an electrically charged population of dust particles can strongly influence the plasma properties, and drive it to behave in a very different manner in comparison with a conventional ion-electron plasma. Another common feature of space plasmas is the presence of Alfv{\'e}n waves. These low-frequency waves are generated by oscillations of the magnetic field and play important roles in several astrophysical mechanisms, such as the heating and transport of energy in stellar winds, transfer of angular momentum in interstellar molecular clouds, and providing scattering mechanisms for the acceleration of cosmic rays \citep[][]{cramer2001physics}.

Numerous theoretical investigations of how these waves might heat and accelerate the solar wind have been proposed \citep[e.g.][]{Barnes1969,Alazraki1971,belcher1971alfvenic,Belcher1971_LargeAlfven,Heinemann1980,jatenco1989effect,Falceta_Goncalves_2002}. This idea is supported by the fact that Alfv{\'e}n waves were detected in the solar wind, firstly trough \emph{in situ} observations by the Mariner V spacecraft \citep[][]{Belcher1971_LargeAlfven}, and more recently by the Ulysses spacecraft near the Earth \citep[][]{Smith_1995Ulysses,Goldstein_1995}. Advances coming from ground-based \citep[][]{Tomczyk_2007Alfven} and space-based instrumentation \citep[][]{dePontieu2007chromospheric} provide further evidence of a significant energy flux near the solar corona caused by Alfv{\'e}n waves. 

It was pointed out by \citet{Leamon1999} that purely parallel propagating Alfv{\'e}n waves are not sufficient to describe the observed characteristics of the solar wind. Some theoretical and observational results suggest that kinetic Alfv{\'e}n waves (KAW), which are essentially highly oblique Alfv{\'e}n waves, must play an important role in particle acceleration \citep[][]{hui1992electron} and in turbulent processes in the solar wind \citep[][]{leamon1998observational,Leamon_2000,Bale2005}. In the kinetic regime, the electron thermal speed is higher than the Alfv{\'e}n speed and the magnetic-aligned electric field produced by obliquely propagating waves is expected to produce strong wave-particle interactions. For this reason, KAWs are often associated with the heating and acceleration of electrons along the plasma magnetic field lines in solar flares and in aurora arcs \citep[][]{hasegawa1976particle, Goertz1979, Bian2020}.

In uniform plasmas, it is well known that Alfv{\'e}n waves consist of the compressional and shear Alfv{\'e}n waves (CAW and SAW) which may also be denominated as, respectively, whistler and ion-cyclotron waves in the large wavenumber regime. For parallel propagation, these two waves present null values of parallel electric field and, for a dustless plasma, they couple into a single mode in the short wavenumber region; however, for oblique propagation, these modes no longer couple \citep[][]{Gaelzer_2008}. The CAW compresses the magnetic field lines, which acts as a restoring force, and propagates isotropically, with its group velocity along the same direction of its phase velocity \citep[][]{chen2021physics}. On the other hand, for SAW, the magnetic field lines twist relative to one another but do not compress. One of the most important properties of SAW is that it is an anisotropic electromagnetic wave, i.e., whilst its phase velocity can propagate in any direction, its group velocity propagates along the magnetic field \citep[][]{swanson2003plasma}. However, \citet{Vasconez2015} have shown that for dispersive Alfv{\'e}n waves all modes have a non-vanishing component of the group velocity in the perpendicular direction to the magnetic field, although for SAW it is much less than the parallel component.
 
A substantial fraction of the Alfv{\'e}n waves energy flux is transferred to plasma particles by some form of dissipation, being Landau damping the usual process for a dustless plasma. However, several works showed, with the aid of the kinetic theory, that the dispersion relation in a magnetized dusty plasma is greatly modified, with the normal modes showing a stronger damping mechanism than the usual Landau damping, caused by the inelastic collisions that occur between the electrically charged dust grains and plasma particles \citep[see e.g.][]{dejuli_2005,deJuli_2007mode,ziebell2008new,Gaelzer_2008,Gaelzer_2010,detoni2021,detoni2022}.

The presence of dust particles near the Solar corona has been observed from infrared emissions, specially during solar eclipses \citep[][]{mankin1974coronal,Lena1974,durst1982two,Leinert1998}, providing evidence of dust particles existing as close as $2\,\text{R}_\odot$, where $\text{R}_\odot$ is the Sun's radius. \emph{In situ} observations by the NASA's Parker Solar Probe \citep[][]{howard2019near} also corroborate the theory that there is a decrease in dust density as it approaches the Sun's surface, since dust grains this close would be heated and vaporised by the intense sunlight. Observations also provide evidence of the existence of dust envelopes around many other stars. Of particular interest is the role that dust particles play in the acceleration mechanism of stellar winds of carbon-rich stars. These stars are notable for losing great amount of their masses by way of these powerful stellar winds \citep[][]{Knapp1987_MassLoss,Lafon1991_MassLoss,Mattson2011}. Several models that describe the mass-loss in late-type stars show that the stellar winds of these stars are usually dust-driven \citep[][]{Bowen1988,Sandin_2003,Woitke_2006,Boulangier_2018}, in which stellar photons, incident on dust particles, will lead to a radiative acceleration of the grains away from the star and, subsequently, momentum will be transferred to the surrounding gas
by gas–grain collisions. In addition to the radiation pressure, these winds are greatly affected by the strong damping of Alfv{\'e}n waves, most likely due to grain interaction \citep[][]{Falceta_Goncalves_2002}.

In view of the importance of Alfv{\'e}n waves in the heating and acceleration of dust-driven stellar winds, a substantial amount of work has been done in order to better comprehend the effects of dust particles in the propagation and damping of these waves, which consider purely parallel propagating waves, with dust particles charged only by the absorption of plasma particles \citep[see e.g.][]{dejuli_2005,Ziebell_2005,deJuli_2007mode,Gaelzer_2010}. \citet{Gaelzer_2008} showed that the dispersion and absorption of obliquely propagating Alfv{\'e}n waves are also substantially modified by the presence of dust, with the ion-cyclotron modes presenting a interval of wavenumber values with null frequency, a feature not observed in a conventional plasma.

Using the formalism developed by \citet{galvao_ziebell2012}, which includes in the kinetic theory of magnetized dusty plasmas the process of photoionization of dust particles, it is possible to observe the effects caused by the dust population in a more realistic way, since the particles in stellar winds are exposed to radiation coming from the star's surface. In this case, the dust population can present positive values for its electric charge, as opposed to the case where they are charged only by absorption of plasma particles, which tends to negatively charge the dust grains.

\citet{detoni2021} first used this formalism to study the changes that the photoionization process may bring to the properties of purely parallel propagating Alfv{\'e}n waves. Some of the results reveal that the coupling between the whistler and ion cyclotron modes is greatly modified in the large wavelength region once dust particles have null or positive electrical charge. Also, it was shown that the presence of photoionization may greatly modify the damping rates seen in both modes. More recently, \citet{detoni2022} extended the study of the effects of dust photoionization to obliquely propagating Alfv{\'e}n waves. It was seen that, for the set of parameters typical of a stellar wind coming from a carbon-rich star, the photoionization process tends to diminish the damping rate of both CAW and SAW modes. Additionally, this charging process reduces the interval of wavenumber values where the SAW present null frequency.

As mentioned before, both CAW and SAW are expected to present non-zero transverse group velocity, which allow the wave energy to propagate across the magnetic field lines. CAW and SAW waves are known to represent an important mode of energy and momentum transport between the tail of Earth's magnetosphere and the auroral zone \citep[][]{keiling2002,Lysak2009}. In this context the magnitude of the non-vanishing perpendicular group velocity can have important consequences for energy transport, since waves moving along the sheet boundary layer in the magnetotail would also move across the magnetic field lines and leave the plasma sheet boundary layer before reaching the inner magnetosphere \citep[][]{malovichko2013properties}.

In nonuniform plasmas in which quantities vary in the direction transverse to the ambient magnetic field causing a large gradient of the Alfv{\'e}n velocity, waves with small amplitude can undergo phase mixing \citep[][]{Heyvaerts1983CoronalHeating}, in which differences in group velocity at different locations progressively bend wave fronts. Also, waves such as SAW, which have group velocity mostly parallel to the magnetic field, can be subject to resonant absorption, in which case the wave energy concentrates in the local field line where its frequency locally matches a characteristic frequency \citep[][]{Chen_1995TheoryofShear}. However, since SAWs have a non-vanishing perpendicular group velocity, the energy propagates across field lines away from the resonant layer; this mechanism can saturate the growth in amplitude of the resonant oscillation and limit the contraction of its transverse size \citep[][]{Bellan1994AlfvenResonance,stasiewicz2000small}. These processes are known to take place in the Earth's magnetosphere and in the Sun's corona, heating and transferring energy to the local plasma \citep[][]{ChenHasegawa1974_PlasmaHeating,HasegawaChen1976_Kinetic}. 

Given the importance and ubiquity of phenomena involving wave propagation and energy transfer in space environments, and taking into account that these are commonly populated by dust particles, in this paper we follow the work of \citet{detoni2022} by studying the characteristics of the group velocity of oblique Alfv{\'e}n waves in a dusty plasma typical of a stellar wind environment. The intention is to better understand how the energy of both CAW and SAW propagates in a magnetized dusty plasma and how their group velocity is changed when varying some of the parameters related to the dust particles, such as number density and equilibrium electrical charge.

The structure of this paper is the following. In Section \ref{sec:the_model}, we discuss the plasma model and the dust charging mechanisms considered. In Section \ref{sec:dispersion_relation}, the dispersion relation for obliquely propagating Alfv{\'e}n waves is derived. Section \ref{sec:results} presents the numerical results and discussions. Finally, the conclusions and final remarks are presented in Section \ref{sec:conclusions}.

\section{The plasma model and dust charging processes} \label{sec:the_model}

The kinetic formalism applied in this work is the same that recently appeared in \citet{detoni2022}, in which is presented a more detailed account of the theory \citep[see also][]{dejuli_schneider_1998,dejuli_2005,Gaelzer_2008,galvao_ziebell2012}. Nevertheless, it is useful to repeat here some of the basic features of the model.


We start by considering a multi-component homogeneous plasma consisting of protons, electrons and spherical dust particles with radius $a$, placed in an external magnetic field $\mathbfit{B}_{0}=B_{0}\mathbfit{e}_{z}$ and exposed to anisotropic incidence of radiation, as it happens in the surrounding of stars. Dust grains embedded in a plasma will acquire their own electrical charge by way of many charging processes, which are represented by currents incident on the grains' surface, i.e.,
\begin{equation}
    \frac{\mathrm{d}q_\mathrm{d}}{\mathrm{d}t} = \sum_k I_k (q_\mathrm{d})\,,
    \label{eq:dustchargevariation}
\end{equation}
where $q_\mathrm{d}$ is the dust electrical charge and $I_k$ is the current correspondent to the $k$-th charging process. After some time, the dust grains will achieve an equilibrium electrical charge $q_\mathrm{d0}$, in which case the total current over the surface will be zero, i.e.,
\begin{equation}
    \frac{\mathrm{d}q_\mathrm{d}}{\mathrm{d}t} \bigg|_{q_d=q_\mathrm{d0}} = \sum_j I_k(q_\mathrm{d0}) = 0\,.
    \label{eq:equilibrium_charge}
\end{equation}

As a consequence of the variable charge of dust particles, the equilibrium number densities $n_{\beta0}$ of the plasma species $\beta$ will change through the quasi-neutrality condition
\begin{equation}
    \sum_{\beta}n_{\beta0}q_{\beta}+q_\mathrm{d0}n_\mathrm{d0}=0,
    \label{eq:quasineutrality}
\end{equation}
where $q_\beta$ is the charge of the plasma species $\beta$, and $n_\mathrm{d0}$ is the equilibrium density of dust particles.

Generally, in both space and laboratory plasmas, more than one charging process may play an important role in the variation of the dust electrical charge. Some important charging mechanisms that occur on dust particles are: the absorption of plasma particles due to inelastic collisions between them and dust grains; electron emission by photoelectric effect; secondary emission of electrons; thermionic emission; among others. In our model, each of these processes would add an additional term in the kinetic equation of plasma particles and greatly complicate the derivation of the dielectric tensor. 

In this work, we consider only the absorption of particles of species $\beta$ and the photoemission of electrons by the dust grains, represented, respectively, by the currents $I_\beta$ and $I_\mathrm{p}$. The absorption current is derived from the orbital motion limited (OML) theory \citep[see e.g.][]{Allen_1992,Tsytovich_1997}, an approximation valid for weakly magnetized plasmas where the dust particle radius is much smaller than the electron Larmor radius \citep[][]{Chang_1993,salimullah2003dust,Kodanova+2019}. For the parameters used in this work, this condition is always satisfied.

The explicit expression of the absorption current in the equilibrium is given by \citep{dejuli_schneider_1998}
\begin{equation}
    I_{\beta0}(q_\mathrm{d})=\pi a^{2}q_{\beta}\int \mathrm{d}^{3}p\left(1-\frac{C_\beta}{p^{2}}\right)H\left(1-\frac{C_\beta}{p^{2}}\right)\frac{p}{m_{\beta}}f_{\beta0},
    \label{eq:I_beta0}
\end{equation}
where $m_\beta$ and $f_{\beta0}$ are, respectively, the mass and the distribution function of the plasma species $\beta$ in equilibrium, $H\left(x\right)$ is the Heaviside function, and
\begin{equation}
    C_\beta \equiv \frac{2 q_\mathrm{d} q_\beta m_\beta }{a}.
\end{equation}

The model for the photoionization current considers that radiation is unidirectional, striking only one side of the dust particle. When the energy of the radiation is greater than the work function of the material, electrons can be emitted. Its expression is given by \citep[see][appendix]{galvao_ziebell2012}
\begin{equation}
    I_\mathrm{p}=\frac{2}{h^{3}}\int \mathrm{d}^{3}p\,\sigma_\mathrm{p}(p,q_\mathrm{d})\frac{p_{z}}{m_{e}}\left[1+\exp\left(\frac{p^{2}}{2m_{e}k_\mathrm{B}T_\mathrm{d}}-\xi\right)\right]^{-1},
    \label{eq:I_p_integral}
\end{equation}
where $h$ is the Planck constant, $T_\mathrm{d}$ is the dust temperature, and
\begin{equation}
    \xi=\frac{1}{k_\mathrm{B}T_\mathrm{d}}(h\nu-\phi)
\end{equation}
with $\nu$ being the incident radiation frequency, and $\phi$ the
work function of the material. The photoemission cross section is written as
\begin{equation}
    \sigma_\mathrm{p}(p,q_\mathrm{d})=e\pi a^{2}\beta(\nu)\Lambda(\nu)H(p_{z})H\left(1-\frac{2m_e e q_d}{a p^{2}}H(q_\mathrm{d})\right),
\end{equation}
where $e$ is the elementary charge, $\beta(\nu)$ is the probability of an electron which arrives to the surface coming from the inside to absorb a photon of frequency $\nu$ at the surface, and $\Lambda(\nu)$ is the number of photons with frequency $\nu$ incident per unit of area per unit of time.

If we consider that the star radiate as a black body, we can express the number of photons with frequency between $\nu$ and $\nu+\mathrm{d}\nu$ incident per unit time per unit area by
\begin{equation}
    \Lambda(\nu)\mathrm{d}\nu=\frac{4\pi\nu^{2}}{c^{2}}\left[\exp\left(\frac{h\nu}{k_\mathrm{B}T_\mathrm{s}}\right)-1\right]^{-1}\left(\frac{r_\mathrm{s}}{r_\mathrm{d}}\right)^{2}\mathrm{d}\nu,
    \label{eq:blackbody}
\end{equation}
where $c$ is the speed of light in vacuum, $T_\mathrm{s}$ is the surface temperature of the star, $r_\mathrm{s}$ is the radius of its radiating surface, and $r_\mathrm{d}$ is the mean distance of the dust grains from the star. 

Both integrals in momenta appearing in equations \eqref{eq:I_beta0} and \eqref{eq:I_p_integral} can be solved. More details on these charging models can be found in \citet{detoni2021} where the absorption current is solved for Maxwellian distribution of plasma particles and the photoemission current is expressed in terms of empirical quantities, such as the maximum photoelectric efficiency $\chi_m$ of the material, and is generalized to the case of a continuous spectrum of the radiation.

As mentioned before, when writing the kinetic equations of the system, these charging processes will alter the equations for plasma particles. On top of that, one must also supply a kinetic equation for the dust population. Therefore, considering that dust particles are motionless and have all the same size, we can write the Vlasov-Maxwell set of equations as \citep[][]{galvao_ziebell2012}
\begin{align}
    &\frac{\partial f_\mathrm{d}}{\partial t} + \frac{\partial}{\partial q} (I f_\mathrm{d})=0 \,, \label{eq:conjVlasovMaxwell1}\\
    &\begin{aligned}
    &\left( \frac{\partial}{\partial t} + \frac{\Vec{p}}{m_\beta} \cdot \nabla_{\Vec{r}} + q_\beta \left[\Vec{E} + \frac{\vec{p}}{m_\beta c} \times \Vec{B} \right] \cdot \nabla_{\vec{p}} \right) f_\beta =  J_\beta + J_\mathrm{p} \,,
    \end{aligned} \label{eq:conjVlasovMaxwell2}\\
    &\nabla \cdot \Vec{E} = 4\pi \sum_\beta q_\beta \int \mathrm{d}^3p f_\beta  + 4\pi \int \mathrm{d} q q f_\mathrm{d}(\vec{r},q,t)  \,, \label{eq:conjVlasovMaxwell3}\\
    &\nabla \cdot \Vec{B} = 0 \,, \label{eq:conjVlasovMaxwell4}\\
    &\nabla \times \Vec{E} = -\frac{1}{c} \frac{\partial \Vec{B}}{\partial t} \,, \label{eq:conjVlasovMaxwell5}\\
    &\nabla \times \Vec{B} = \frac{1}{c} \frac{\partial \Vec{E}}{\partial t} + \frac{4\pi}{c} \sum_\beta \frac{q_\beta}{m_\beta} \int \mathrm{d}^3p \,\vec{p} f_\beta \,, \label{eq:conjVlasovMaxwell6}
\end{align}
where $I$ is the total current incident on dust grains, whilst $f_\mathrm{d}(\vec{r},q,t)$ and $f_\beta(\vec{r},\vec{p},t)$ are, respectively, the distribution functions of dust and plasma particles of species $\beta$. The collisional integral $J_\beta$ is closely related to the $I_\beta$ current and acts as a sink of ions and electrons, since they are absorbed by dust grains. On the other hand, the term $J_\mathrm{p}$ describes a source of electrons that are emitted by the photoionization of dust particles and is related to the $I_\mathrm{p}$ current.

The postulation of immobile dust grains greatly simplifies the kinetic equation for the dust population since it excludes the momentum dependence of $f_\mathrm{d}$. However, as a consequence, it restricts the model to the range of wave frequencies much higher than the characteristic dust frequencies. That is, we must consider the regime in which $\omega\gg\max\left(\omega_\mathrm{d},|\Omega_\mathrm{d}|\right)$, where $\omega_\mathrm{d}$ and $\Omega_\mathrm{d}$ are, respectively, the plasma and cyclotron frequencies of the dust particles. Hence, this model is not adequate to study wave modes that arise from dust dynamics, such as the dust acoustic waves, dust ion acoustic waves, and electrostatic dust ion cyclotron waves.

\section{The dielectric tensor and dispersion relation for oblique Alfv{\'e}n waves} \label{sec:dispersion_relation}

Performing a linear approximation in the set of equations \eqref{eq:conjVlasovMaxwell1}--\eqref{eq:conjVlasovMaxwell6}, where we assume that the fields and distribution functions can be written as a summation of an equilibrium part and a small perturbation, e.g., $f_\beta=f_{\beta0}+f_{\beta1}$, and using the Fourier transform of the perturbation, the dielectric tensor will have the form \citep[][]{galvao_ziebell2012}
\begin{equation}
    \epsilon_{ij}=\epsilon_{ij}^\mathrm{C}+\epsilon_{ij}^\mathrm{A}+\epsilon_{ij}^\mathrm{P},
    \label{eq:dielectric_tensor}
\end{equation}
where $\epsilon_{ij}^\mathrm{C}$ indicates the components which have almost the same formal structure of a conventional (dustless) plasma. The terms $\epsilon_{ij}^\mathrm{A}$ and $\epsilon_{ij}^\mathrm{P}$ are related to the dust charging processes considered in the formalism, and would not appear in a dustless plasma. The former appears due to the absorption of plasma particles by the dust, whilst the latter is related to the photoionization of dust particles. As a first approach to the problem, we consider only the `conventional' part of the dielectric tensor, i.e., we make $\epsilon_{ij}=\epsilon_{ij}^\mathrm{C}$.

Therefore, for a Maxwellian distribution of plasma particles, the dielectric tensor can be written as
\begin{equation}
\begin{aligned}
    \epsilon_{ij} = \delta_{ij} + \sum_{n=-\infty}^{\infty} \sum_{\beta} \frac{\omega_{p\beta}^2}{\omega n_{\beta0}} \int \mathrm{d}^{3}p     \left(\frac{p_\parallel}{p_\perp} \right)^{\delta_{jz}} \frac{\partial f_{\beta0}}{\partial p_{\perp}} \\
    \times \frac{p_{\parallel}^{\delta_{iz}} p_{\perp}^{\delta_{ix}+\delta_{iy}} \Pi_{ij}^{n\beta}}{\omega-n\Omega_\beta-\frac{k_{\parallel}p_{\parallel}}{m_{\beta}}+\mathrm{i}\nu_{\beta d}^{0}(p)},
\end{aligned}
\label{eq:tensor_conventional}
\end{equation}
where $\omega$ is the angular frequency, $\omega_{p\beta}$ and $\Omega_\beta$ are, respectively, the plasma and cyclotron frequencies of particles of species $\beta$. The tensor $\Pi_{ij}^{n\beta}$ is related to the Bessel functions and its first derivatives \citep[see e.g.][appendix]{dejuli_2005}.

The presence of dust particles introduce a new imaginary term in the resonant denominator of equation \eqref{eq:tensor_conventional}, featuring the inelastic collision frequency $\nu_{\beta d}^{0}(p)$ between dust and plasma particles. To evaluate the momentum integral, we replace this term by its average value in momentum space
\begin{equation}
    \nu_{\beta}=\frac{1}{n_{\beta0}}\int \mathrm{d}^{3}p\,\nu_{\beta d}^{0}(p)f_{\beta0}.
\end{equation}
This procedure enable us to derive relatively simple expressions for the dispersion relation and it has been shown to be a good approximation in the frequency range of Alfv{\'e}n waves \citep[][]{deJuli_2007mode}.

The dispersion relation for a homogeneous magnetized plasma with $\mathbfit{B}_{0}=B_{0}\mathbfit{e}_{z}$ and a wave vector $\mathbfit{k}$ lying on the $xz$ plane is given by
\begin{equation}
    \det\begin{pmatrix}\epsilon_{xx}-N_{\text{\ensuremath{\parallel}}}^{2} & \epsilon_{xy} & \epsilon_{xz}+N_{\text{\ensuremath{\parallel}}}N_{\text{\ensuremath{\perp}}}\\
    \epsilon_{yx} & \epsilon_{yy}-N^{2} & \epsilon_{yz}\\
    \epsilon_{zx}+N_{\text{\ensuremath{\parallel}}}N_{\text{\ensuremath{\perp}}} & \epsilon_{zy} & \epsilon_{zz}-N_{\text{\ensuremath{\perp}}}^{2}
    \end{pmatrix}=0,
    \label{eq:Dispersion-equation}
\end{equation}
where $\mathbfit{N}=N_{\perp}\mathbfit{e}_{x}+N_{\parallel}\mathbfit{e}_{z}=\mathbfit{k}c/\omega$ is the refractive index.

We now introduce the following dimensionless parameters
\begin{equation}
\begin{alignedat}{2}
    &z=\frac{\omega}{\Omega_{i}},\quad \varepsilon=\frac{n_{d0}}{n_{i0}},\quad  u_{\beta}=\frac{v_{T\!\beta}}{v_\mathrm{A}},\quad \chi_{\beta}=\frac{q_\mathrm{d0}q_{\beta}}{ak_\mathrm{B}T_{\beta}},\\
    &\gamma=\frac{\lambda^{2}n_{i0}v_\mathrm{A}}{\Omega_{i}},\quad  \tilde{a}=\frac{a}{\lambda},\quad \lambda=\frac{e^{2}}{k_\mathrm{B}T_{i}},\quad  \mathbfit{q}=\frac{\mathbfit{k}v_\mathrm{A}}{\Omega_{i}},\\
    &\tilde{\nu}_{\beta}=\frac{\nu_{\beta}}{\Omega_{i}},\quad \eta_{\beta}=\frac{\omega_{p\beta}}{\Omega_{i}},\quad  r_{\beta}=\frac{\Omega_{\beta}}{\Omega_{i}},
\end{alignedat}
\end{equation}
where $v_\mathrm{A}$ is the Alfv{\'e}n velocity,
\begin{equation}
    v_\mathrm{A}^{2}=\frac{B_{0}^{2}}{4\pi n_{i0}m_{i}}.
\end{equation}

The group velocity, given the cylindrical symmetry and that $\mathbfit{k}$ is lying on the $xz$ plane, is expressed as
\begin{equation}
    \mathbfit{v}_g = \frac{\partial \omega}{\partial k_\perp} \mathbfit{e}_x +  \frac{\partial \omega}{\partial k_\parallel} \mathbfit{e}_z \,,
\end{equation}
and, in terms of these dimensionless quantities, can be written as
\begin{equation}
    \mathbfit{u}_g = \frac{\partial z}{\partial \mathbfit{q}} = \frac{\mathbfit{v}_g}{v_\mathrm{A}}\,.
    \label{eq:u_g_derivative}
\end{equation}

Considering that the studied modes have large wavelength in the perpendicular direction, i.e., $q_\perp\ll1$, and keeping only the $n=-1,0,1$ harmonics in the dielectric components, the dispersion relation for obliquely propagating waves can be written as follows,
\begin{equation}
    \begin{aligned}
        &\bigg[\bigg(\frac{z^{2}}{\eta_{i}^{2}}+\epsilon_{yy}^{1}-q_{\parallel}^{2}\bigg) \bigg(\frac{z^{2}}{\eta_{i}^{2}}+\epsilon_{xx}^{1}-q_{\parallel}^{2}\bigg) - (\epsilon_{xy}^{1})^{2} \bigg] \bigg(\frac{z^{2}}{\eta_{i}^{2}}+\epsilon_{zz}^{0}\bigg)\\
        &+\Bigg\{\bigg(\frac{z^{2}}{\eta_{i}^{2}}+\epsilon_{zz}^{0}\bigg) \bigg(\frac{z^{2}}{\eta_{i}^{2}}+\epsilon_{xx}^{1}-q_{\parallel}^{2}\bigg) \bigg(\epsilon_{yy}^{0}-1\bigg)\\
        &\quad+\bigg(\frac{z^{2}}{\eta_{i}^{2}}+\epsilon_{yy}^{1}-q_{\parallel}^{2}\bigg) \bigg[ \bigg(\epsilon_{zz}^{1}-1\bigg) \bigg(\frac{z^{2}}{\eta_{i}^{2}}+\epsilon_{xx}^{1}-q_{\parallel}^{2}\bigg) \\
        &\quad- \bigg(\epsilon_{xz}^{1}+q_{\parallel}\bigg)^{2} \bigg] -\bigg[ - 2\epsilon_{xy}^{1}\epsilon_{yz}^{1}\bigg(\epsilon_{xz}^{1}+q_{\parallel}\bigg) \\
        &\quad+(\epsilon_{xy}^{1})^{2}\bigg(\epsilon_{zz}^{1}-1\bigg)+(\epsilon_{yz}^{1})^{2}\bigg(\frac{z^{2}}{\eta_{i}^{2}}+\epsilon_{xx}^{1}-q_{\parallel}^{2}\bigg) \bigg] \Bigg\} q_{\perp}^{2} \\
        &+\Bigg\{\bigg(\epsilon_{yy}^{0}-1\bigg)\bigg[\bigg(\frac{z^{2}}{\eta_{i}^{2}}+\epsilon_{xx}^{1}-q_{\parallel}^{2}\bigg)\bigg(\epsilon_{zz}^{1}-1\bigg)\\
        &\quad-\bigg(\epsilon_{xz}^{1}+q_{\parallel}\bigg)^{2}\bigg]\Bigg\} q_{\perp}^{4} =0,
    \end{aligned}
    \label{eq:disp_rel}
\end{equation}
where $\eta_i=c/v_\mathrm{A}$ and the $\epsilon_{ij}^{0,1}$ tensor components are related to the $\epsilon_{ij}$ components. A more detailed account of this derivation and explicit expressions for these components can be found in \citet{Gaelzer_2008}.

\section{Numerical Results} \label{sec:results}
To solve the dispersion relation we consider the same set of parameters used in previous works to study Alfv{\'e}n waves within the kinetic theory \citep[e.g.][]{dejuli_2005,Ziebell_2005,Gaelzer_2008,Gaelzer_2010,detoni2021,detoni2022}, i.e., $B_{0}=1$\,G, $n_{i0}=10^{9}$\,cm$^{-3}$, $T_{i}=10^4$\,K, $T_{e}=T_{i}$ and $a=10^{-4}$\,cm. This set of parameters corresponds to values typically found in the winds of carbon-rich stars \citep[][]{tsytovich_2004}.

In this environment, dust particles are composed mainly of carbon \citep[][]{nanni2021dust}, for that reason, we choose the following values for the dust's physical parameters: work function $\phi=4.6$\,eV; maximum photoelectric efficiency $\chi_\mathrm{m}=0.05$; and dust temperature $T_\mathrm{d}=300$\,K. The chosen value of $T_\mathrm{d}$ is within the range of dust temperatures in the inner circumstellar dust shells of carbon stars \citep{gail_sedlmayr_2014}.

To calculate the radiation flux from equation \eqref{eq:blackbody} we consider a distance of $r_\mathrm{d}=2\,r_\mathrm{s}$, which is within the region of dust formation in dust-driven winds of carbon stars \citep[][]{Danchi1994,gail_sedlmayr_2014}. Since we are dealing with the inner dust shells, we consider that there is no dust between the font and the studied region, so we may neglect the reddening effect of radiation that could occur as a result of the extinction of light caused by other dust particles.

The surface temperatures of carbon stars may vary widely from around $T_s=2000$\,K to over $5000\,$K \citep[][]{wallerstein_1998_carbonstars}. For the parameters used in this work, stars with temperatures above $3500\,$K start presenting significant effects on Alfv{\'e}n waves caused by photoionization of dust particles, so that we choose to work with temperatures above this value in our numerical analysis.

\begin{figure}
    \centering
    \includegraphics[width=\columnwidth]{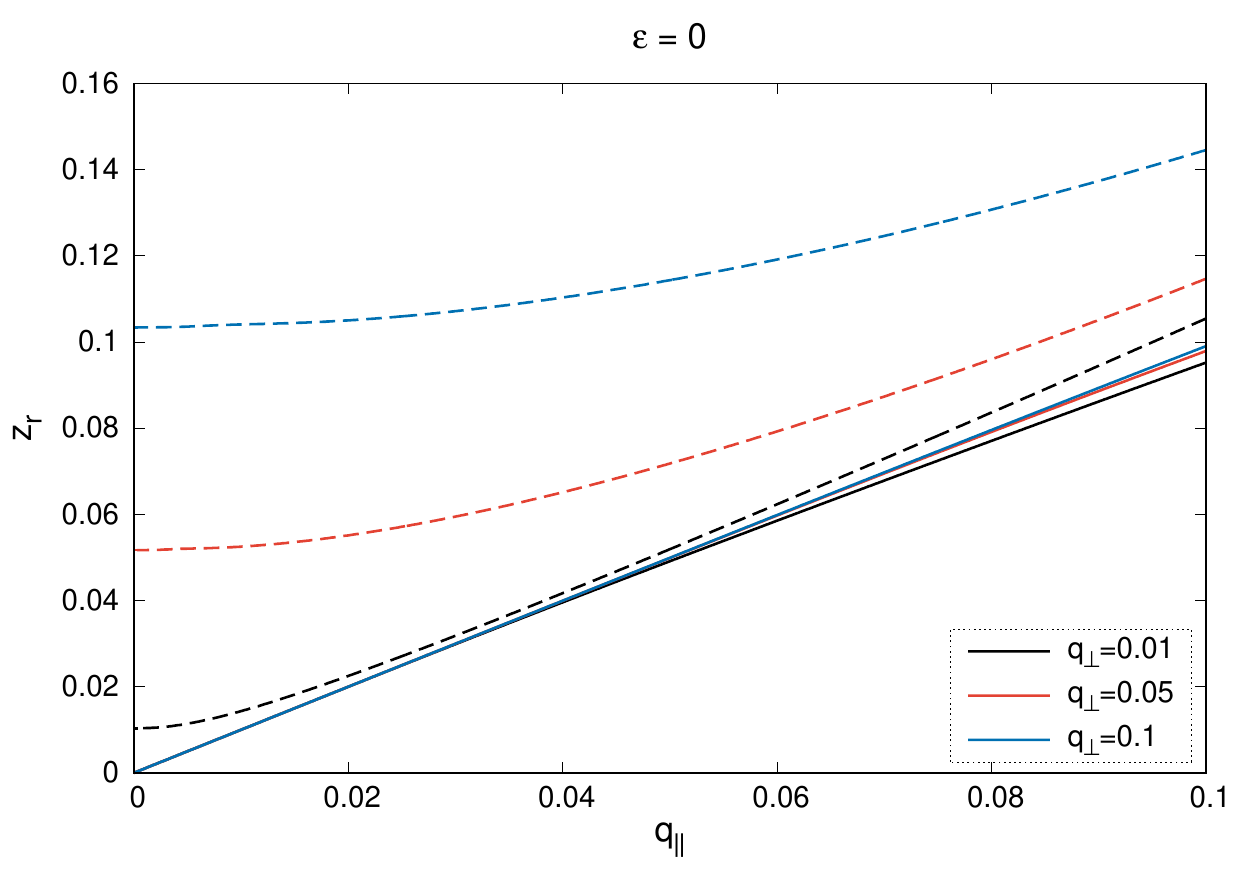}
    \caption{Real part of the normalized frequency $z_\mathrm{r}$ as a function of normalized wavenumber in the parallel direction $q_\parallel$ for several values of perpendicular wavenumber $q_\perp$ for a dustless plasma ($\varepsilon=0$). Continuous lines correspond to SAW whilst dashed lines correspond to the CAW.}
    \label{fig:zr_q_eps0}
\end{figure}

We start our analysis by investigating the group velocity behaviour in a dustless plasma ($\varepsilon=0$), solving the equation~\eqref{eq:disp_rel} using the parameters mentioned above. Fig.~\ref{fig:zr_q_eps0} shows the evolution of the real part of the normalized wave frequency $z_r$ as a function of the parallel normalized wavenumber $q_\parallel$. The perpendicular wavenumber is considered a fixed parameter with values $q_\perp=0.01$, $0.05$ and $0.1$, distinguished by the different line colours. 

This figure is useful to identify the different wave modes that appear from the dispersion relation. Continuous lines represent the SAW for small wavenumber and become the ion cyclotron mode for large wavenumber. We can see that for small values of $q_\parallel$, this mode is not sensitive to the chosen value of $q_\perp$. It is only for larger $q_\parallel$ that is possible to see some small differences between the different line colours for the ion cyclotron mode.

On the other hand, the dashed lines present a greater sensitivity to changes in $q_\perp$. These lines identify the CAW for small wavenumber and become the whistler mode for large wavenumber. We notice that for very small $q_\parallel$, this mode approaches a non-zero value of wave frequency $z_r$, which decreases together with the $q_\perp$ parameter. For purely parallel propagation, the whistler wave approaches $z_r=0$ for small wavenumber and couple with the ion cyclotron mode into a single wave mode in a dustless plasma \citep[see e.g.][]{dejuli_2005,detoni2021}.

\begin{figure}
    \centering
    \begin{minipage}[c]{\columnwidth}
        \centering
        \includegraphics[width=\columnwidth]{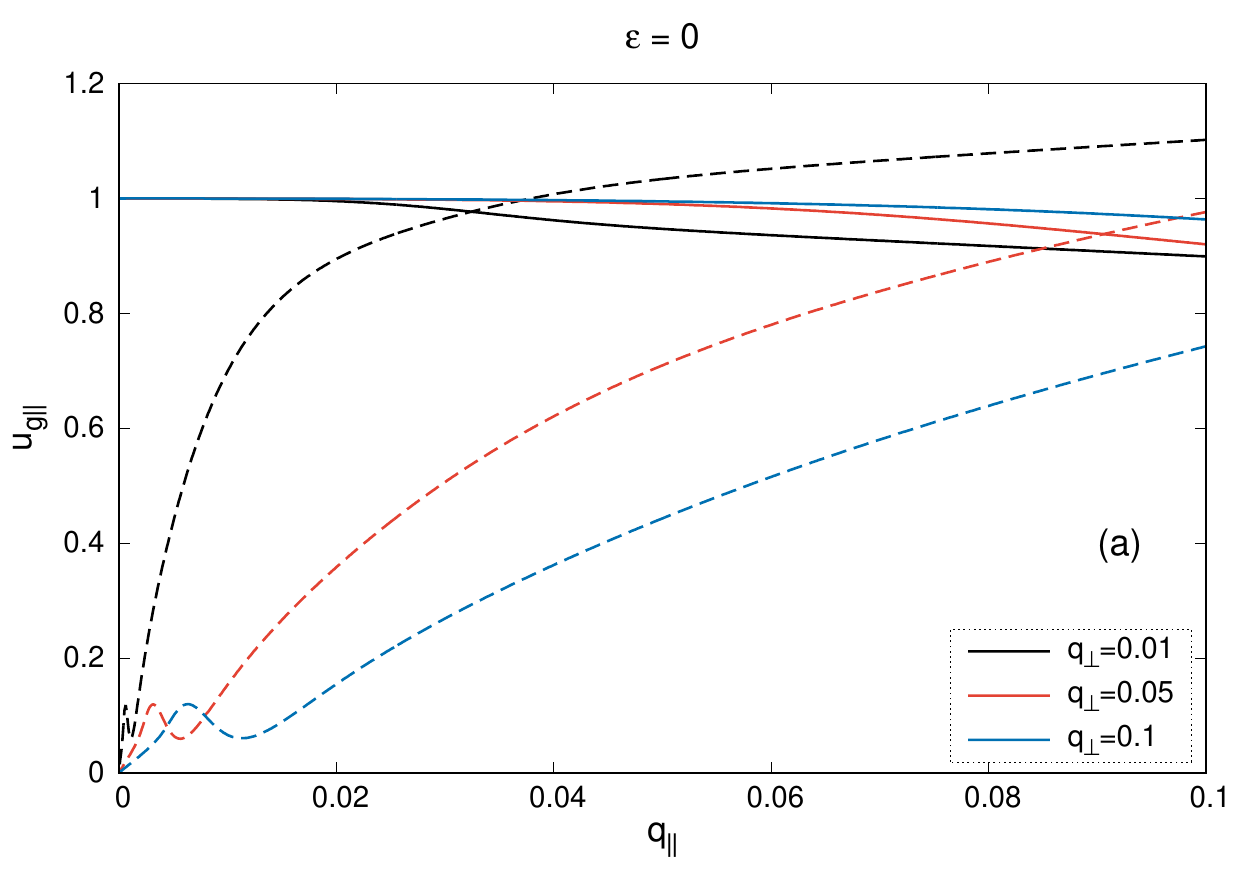}%
    \end{minipage}\vspace{\floatsep}
    \begin{minipage}[c]{\columnwidth}
        \centering
        \includegraphics[width=\columnwidth]{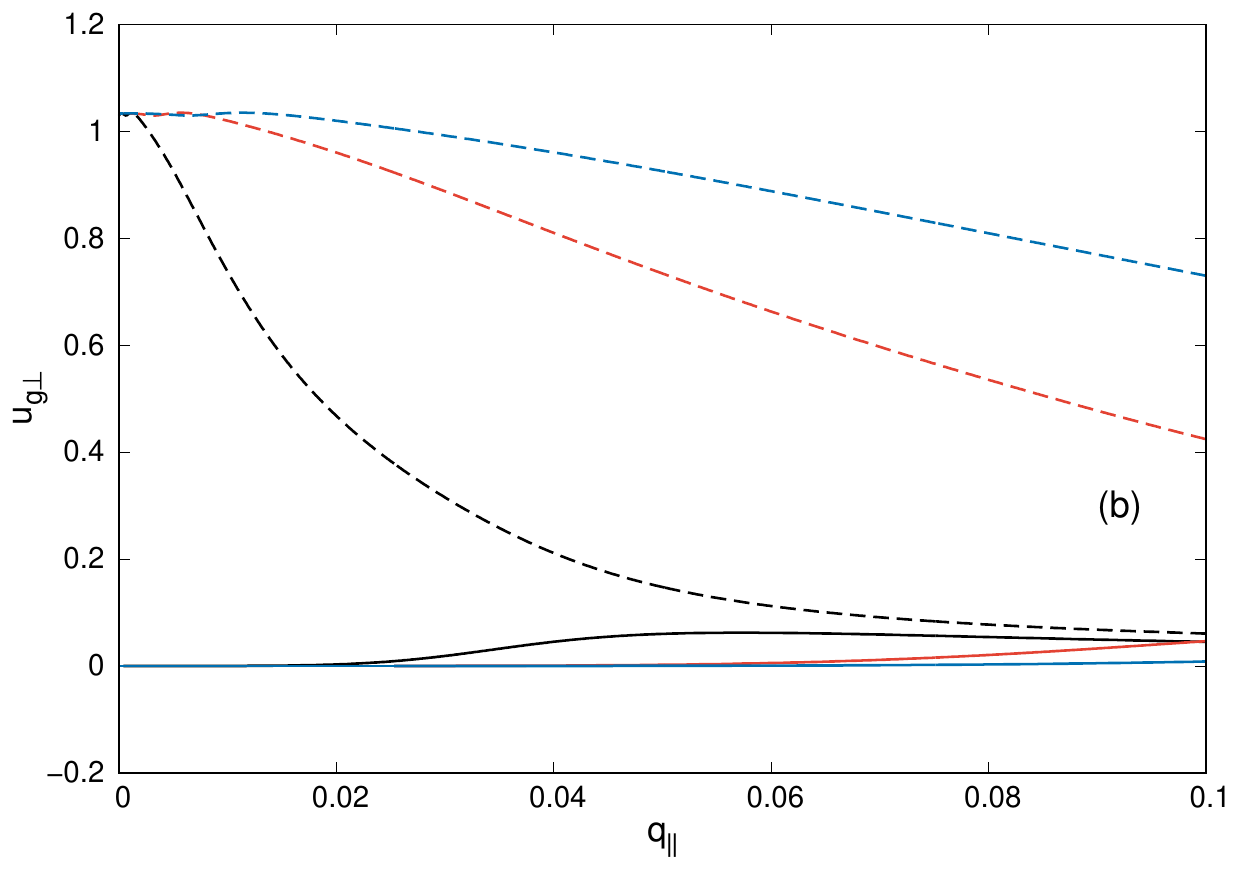}
    \end{minipage}
    \caption{(a) Parallel and (b) perpendicular component of normalized group velocity $u_g$ as functions of $q_\parallel$ for the wave modes presented in Fig.~\ref{fig:zr_q_eps0}. Continuous lines are the SAW/ion cyclotron waves whilst dashed lines are the CAW/whistler waves.}
    \label{fig:groupvel_eps0}
\end{figure}

Now we look at the group velocity of these modes from the numerical differentiation of the dispersion relation. This is done by a simple two-point estimation
\begin{equation}
    \frac{f(x+h)-f(x)}{h} \,,
    \label{eq:numerical_derivative}
\end{equation}
which gives the derivative of the function $f(x)$ in the limit of $h$ approaching zero. However, from the computational point of view, the finite difference approximation of the derivative suffers from catastrophic cancellation for very small $h$, which restricts the practical use of the formula for $h$ smaller than a critical value, corresponding to the greatest obtainable accuracy for the approximation.

Fig.~\ref{fig:groupvel_eps0} shows the parallel (top panel) and perpendicular (bottom panel) components of $\mathbfit{u}_g=\mathbfit{v}_g/v_\mathrm{A}$, i.e., the derivative given by equation \eqref{eq:u_g_derivative}, which gives information about the velocity of a wave packet and energy propagation along and perpendicular to the magnetic field lines, for the modes presented in Fig.~\ref{fig:zr_q_eps0}. 

We see that the SAW/ion-cyclotron waves (continuous lines) present parallel group velocity similar to the Alfv{\'e}n velocity for small $q_\parallel$, whilst the perpendicular component is near zero in that region, regardless of their $q_\perp$ value. The group velocity of these modes only change slightly as the waves present higher values of $q_\parallel$.

However, the group velocity of CAW/whistler waves (dashed lines) displays larger changes as the wavenumber values are modified. The parallel component of group velocity presents higher values for increasing $q_\parallel$, with the exception of a short interval in the small $q_\parallel$ region in which $u_{g\parallel}$ decreases. The component $u_{g\perp}$, on the other hand, tends to decrease as $q_\parallel$ increases, more sharply for smaller values of $q_\perp$. 

Furthermore, we notice that the values of $u_{g\perp}$ for SAW are much lower than $u_{g\parallel}$, which will also happen in the subsequent studies in this work (see Fig.~\ref{fig:zr_q_eps_ioncyc} and \ref{fig:zr_q_Ts_ioncyc}). This indicates that the wave energy is transported predominantly along the magnetic field lines, which is expected for this wave mode \citep[][]{chen2014physics,chen2021physics}.

It is worth mentioning that Figs.~\ref{fig:zr_q_eps0} and \ref{fig:groupvel_eps0} present only the positive roots of the dispersion relation, which correspond to forward-propagating waves relative to the magnetic field. There are also negative solutions \citep[see e.g.][]{Gaelzer_2008,detoni2022}, representing backward-propagating waves, with the same absolute values of $z_r$ and $u_g$, but opposite signs. For the sake of clarity, we present only the solutions for forward-propagating waves throughout this work, keeping in mind that there are also solutions to the dispersion relation which are symmetric about the x-axis.

\begin{table}
	\centering
	\caption{Numerical values for the dust charge number $Z_\mathrm{d}$ for the parameters given in the beginning of this section. The left half of the table shows the cases with constant stellar surface temperature $T_s = 3500\,$K and several values of dust to ion density ratio $\varepsilon$, as studied in Fig.~\ref{fig:zr_q_Ts_eps}(a). The right side corresponds to the curves in Fig.~\ref{fig:zr_q_Ts_eps}(b), with constant $\varepsilon=5\times10^{-6}$ and varying $T_s$.}
	\label{tab:Zd}
	\begin{tabular}{ccccc} 
	    \hline
	    \multicolumn{2}{|c|}{$T_s = 3500\,$K} &\vline& \multicolumn{2}{|c|}{$\varepsilon=5\times10^{-6}$}\\
		\hline 
		$\varepsilon$ & $Z_\mathrm{d}$ &\vline& $T_s (\text{K})$ & $Z_\mathrm{d}$\\
		\hline
		$10^{-7}$          & $-1431$ & \vline & $0$    & $-1495$\\
		$10^{-6}$          & $-1430$ & \vline & $3500$ & $-1427$\\
		$2.5\times10^{-6}$ & $-1429$ & \vline & $4000$ & $-1025$\\
		$5.0\times10^{-6}$ & $-1427$ & \vline & $4500$ & $-205$\\
		$7.5\times10^{-6}$ & $-1425$ & \vline & $5000$ & $+349$\\
		$10^{-5}$          & $-1424$ & \vline & $5500$ & $+699$\\
		\hline
	\end{tabular}
\end{table}

Now we look at the effects of dust particles on these wave modes. First of all, we observe that the dust particles will acquire an equilibrium electrical charge $q_{d0}=Z_\mathrm{d} e$, where $Z_d$ is the dust charge number, which will depend on the plasma parameters. Table~\ref{tab:Zd} presents the values of the dust charge number utilising the set of parameters previously discussed for several values of dust to ion density $\varepsilon$ and radiation flux, given by the stellar surface temperature $T_s$.

We notice that by fixing $T_s$ and varying $\varepsilon$, the charge number is almost not affected at all for the values considered. Nonetheless, as we will see, the wave modes and its group velocities may present significant changes in their behaviour when we increase the dust density. This is due to the inelastic collision frequency that appears in the resonant denominator of equation \eqref{eq:tensor_conventional}, which is affected by this parameter.

On the other hand, the dust charge number in Table~\ref{tab:Zd} is significantly altered by varying $T_s$, to the point where dust particles become positively charged, a consequence of the photoionization process. In this case, the inelastic collision frequency will be altered specially by the changes in the equilibrium dust charge and in the electron density, since the changes in the dust electrical charge will also modify the plasma particles densities by the quasi-neutrality condition (see equation \eqref{eq:quasineutrality}).

\begin{figure}
    \centering
    \begin{minipage}[c]{\columnwidth}
        \centering
        \includegraphics[width=\columnwidth]{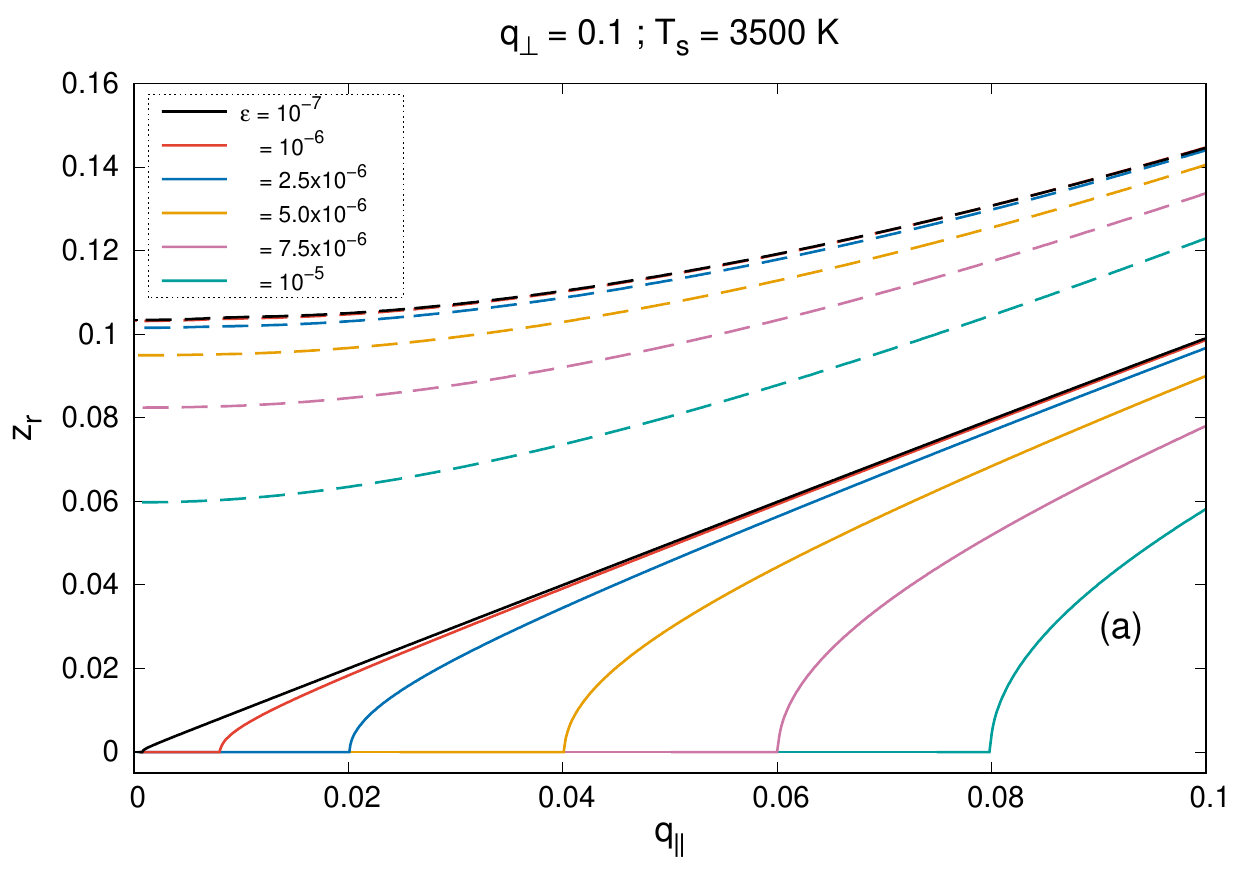}%
    \end{minipage}\vspace{\floatsep}
    \begin{minipage}[c]{\columnwidth}
        \centering
        \includegraphics[width=\columnwidth]{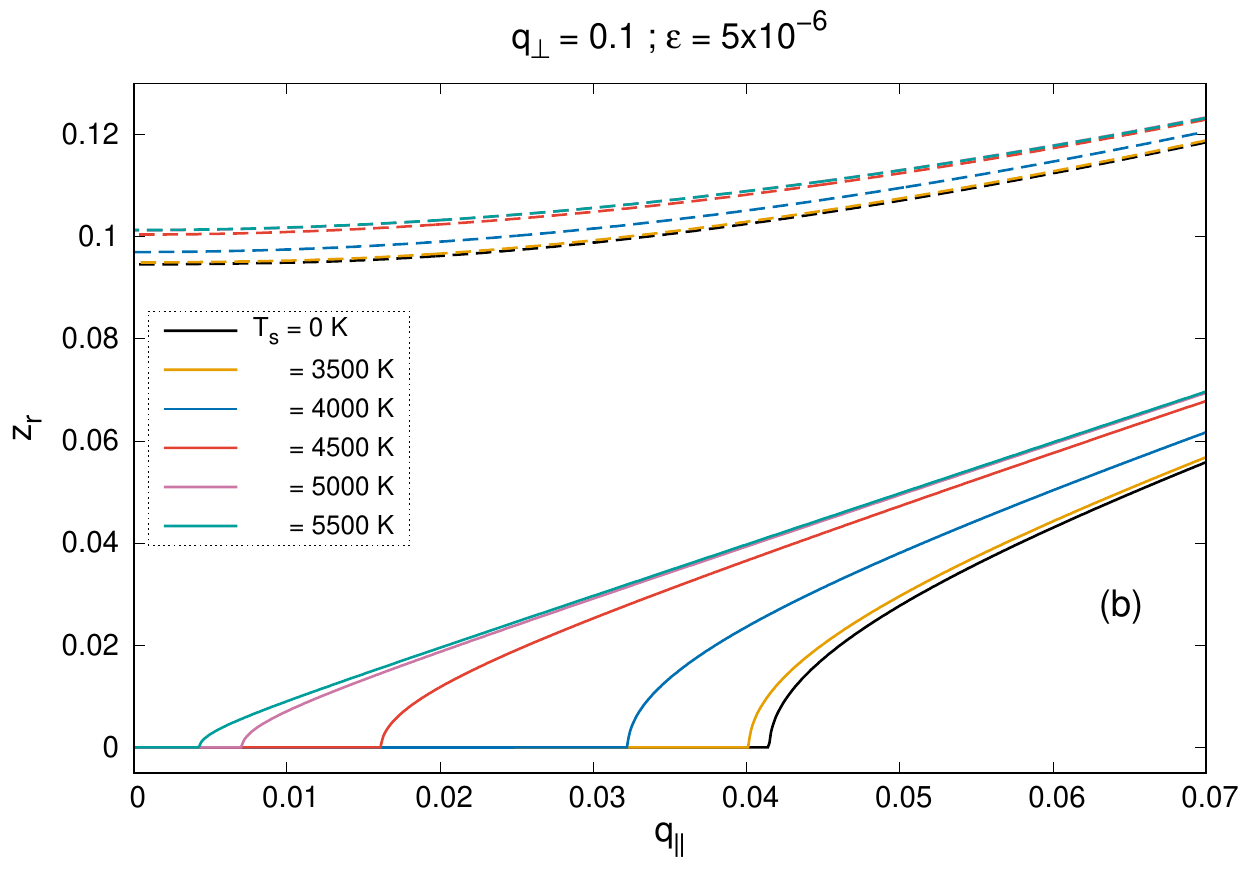}
    \end{minipage}
    \caption{Real part of the normalized frequency $z_r$ as a function of normalized parallel wavenumber $q_\parallel$ and fixed $q_\perp=0.1$ for: (a) fixed value of stellar surface temperature $T_s=3500\,$K and several values of dust to ion density ratio $\varepsilon$; (b) fixed $\varepsilon=5\times10^{-6}$ and varying the value of $T_s$. Dashed and continuous lines represent, respectively, the forward-propagating CAW/whistler modes and the SAW/ion-cyclotron modes.}
    \label{fig:zr_q_Ts_eps}
\end{figure}

Fig.~\ref{fig:zr_q_Ts_eps} displays the forward-propagating wave modes that appear for the set of parameters previously discussed. As before, the dashed and continuous lines represent, respectively, the CAW and SAW. By fixing the value of $q_\perp=0.1$, we observe that when the dust to ion density ratio $\varepsilon$ is not zero, the SAWs present a region of null $z_r$. This non-propagating region appears in the presence of dust particles and was already studied in the works of \citet{Gaelzer_2008} and \citet{detoni2022}. The results show that the interval of $q_\parallel$ where the real frequency is zero is decreased for smaller dust density (top panel) and for higher radiation flux (bottom panel).

\begin{figure}
    \centering
    \begin{minipage}[c]{\columnwidth}
        \centering
        \includegraphics[width=\columnwidth]{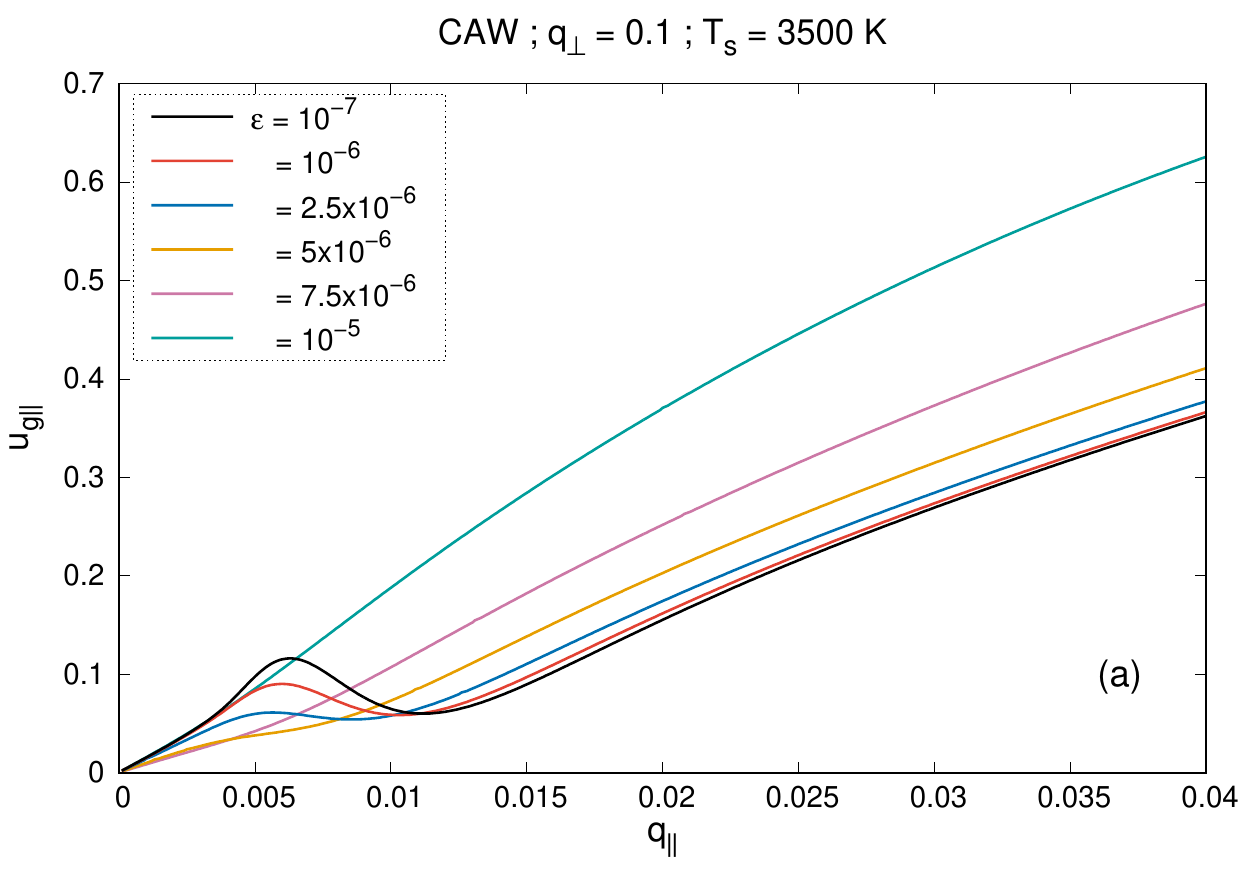}%
    \end{minipage}\vspace{\floatsep}
    \begin{minipage}[c]{\columnwidth}
        \centering
        \includegraphics[width=\columnwidth]{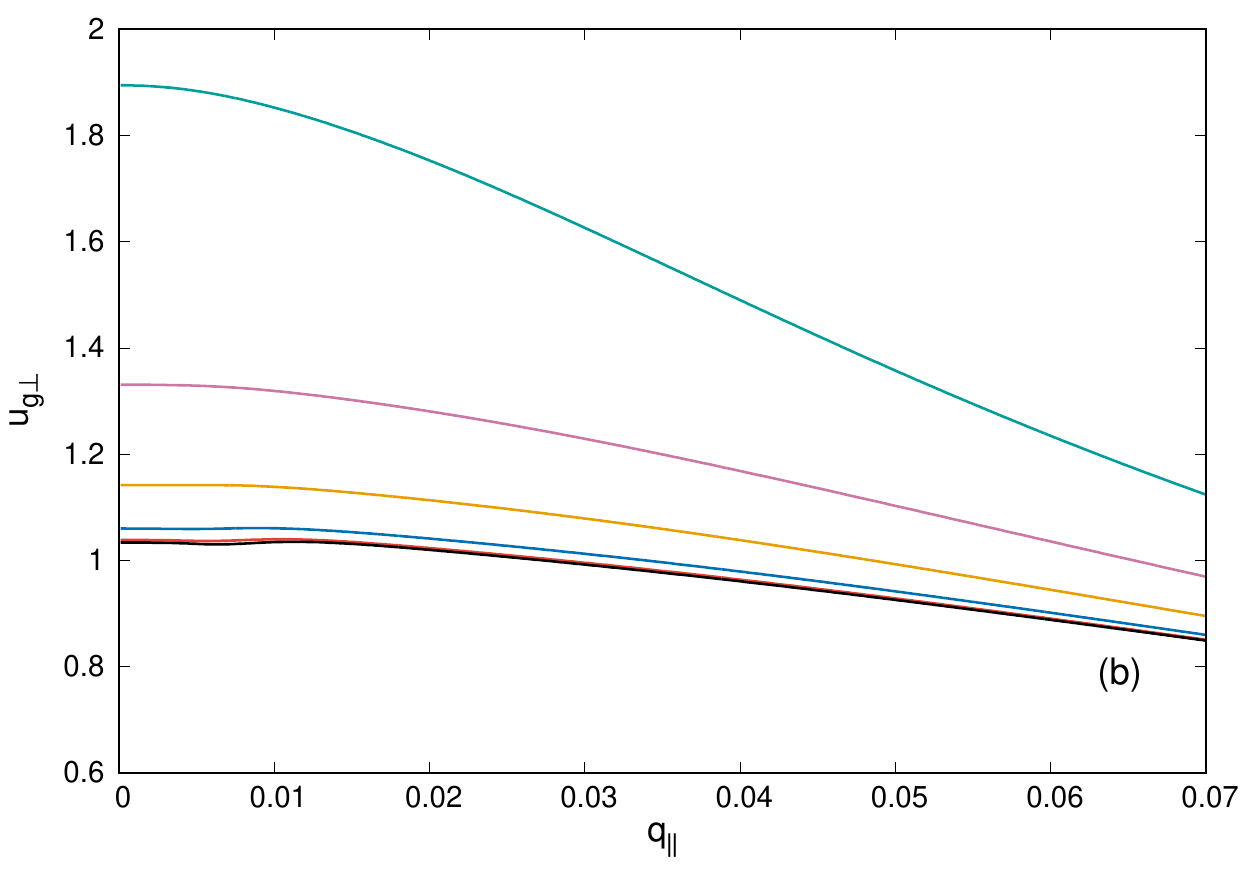}
    \end{minipage}
    \caption{(a) Parallel and (b) perpendicular component of normalized group velocity $u_g$ as functions of $q_\parallel$ for the CAWs presented in Fig.~\ref{fig:zr_q_Ts_eps}(a), with constant stellar surface temperature $T_s=3500\,$K and several values of dust to ion density ratio $\varepsilon$.}
    \label{fig:zr_q_eps_whistler}
\end{figure}

\begin{figure}
    \centering
    \begin{minipage}[c]{\columnwidth}
        \centering
        \includegraphics[width=\columnwidth]{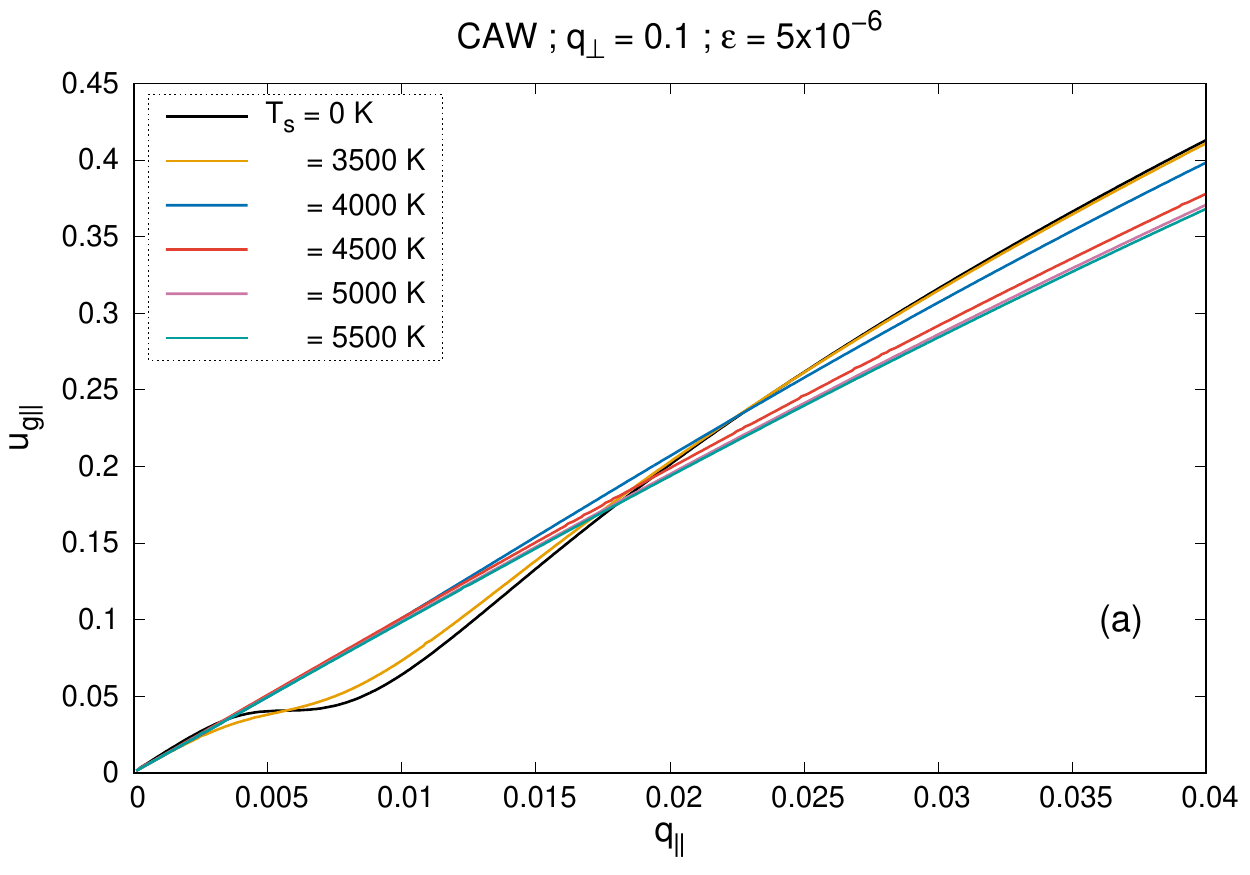}%
    \end{minipage}\vspace{\floatsep}
    \begin{minipage}[c]{\columnwidth}
        \centering
        \includegraphics[width=\columnwidth]{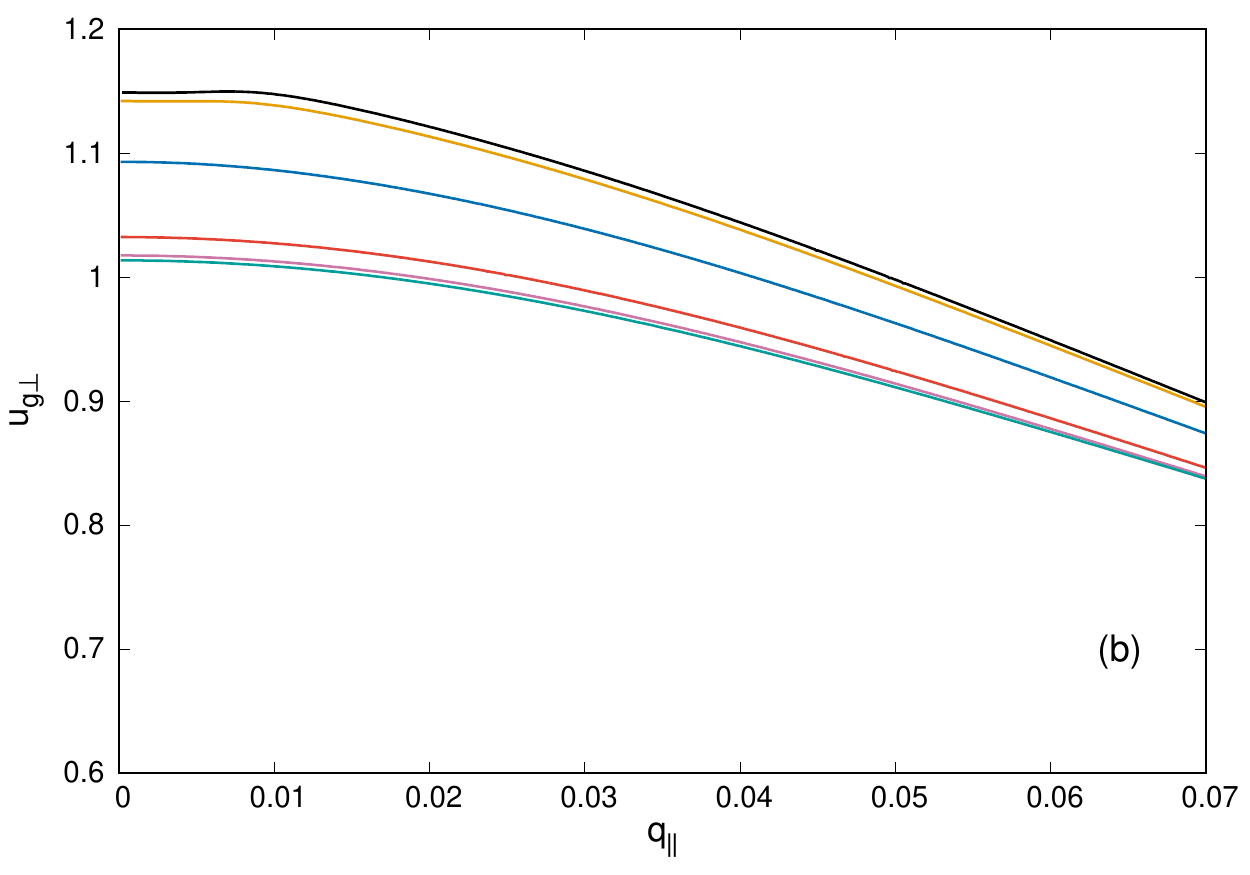}
    \end{minipage}
    \caption{(a) Parallel and (b) perpendicular component of normalized group velocity $u_g$ as functions of $q_\parallel$ for the CAWs presented in Fig.~\ref{fig:zr_q_Ts_eps}(b), with constant dust to ion density ratio $\varepsilon=5\times10^{-6}$ and several values of stellar surface temperature $T_s$.}
    \label{fig:zr_q_Ts_whistler}
\end{figure}

We now look to the group velocities of each mode separately in both cases, varying dust density and varying the radiation flux. Figs.~\ref{fig:zr_q_eps_whistler} and \ref{fig:zr_q_Ts_whistler} show the parallel (top panels) and perpendicular (bottom panels) components of the group velocity of the CAW/whistler modes presented in Fig.~\ref{fig:zr_q_Ts_eps}. We notice that the behaviour of the group velocity for small dust density is very similar to what is observed for a dustless plasma, in Fig.~\ref{fig:groupvel_eps0}, with $u_{g\parallel}$ still presenting decreasing values for a short interval of $q_\parallel$ values (around $0.005 \lesssim q_\parallel \lesssim 0.01$). However, Figs.~\ref{fig:zr_q_eps_whistler}(a) and \ref{fig:zr_q_Ts_whistler}(a) show that this feature tends to disappear for higher values of dust density or radiation flux, with $u_{g\parallel}$ monotonically increasing in all interval of $q_\parallel$ studied.

The perpendicular component $u_{g\perp}$ of CAWs also shows little changes for small dust density when compared to the dustless plasma case. Figs.~\ref{fig:zr_q_eps_whistler}(b) and \ref{fig:zr_q_Ts_whistler}(b) show that the values of $u_{g\perp}$ tends to increase in cases with higher values of $\varepsilon$ or smaller values of $T_s$. For instance, a dust to ion density ratio of $\varepsilon=10^{-5}$ may cause the perpendicular group velocity of waves with small $q_\parallel$ to achieve values almost two times bigger than the Alfv{\'e}n velocity, for the parameters considered, indicating that the propagation of the wave packet for small $q_\parallel$ will occur at greater angles from the magnetic field for higher dust density.

\begin{figure}
    \centering
    \begin{minipage}[c]{\columnwidth}
        \centering
        \includegraphics[width=.99\columnwidth]{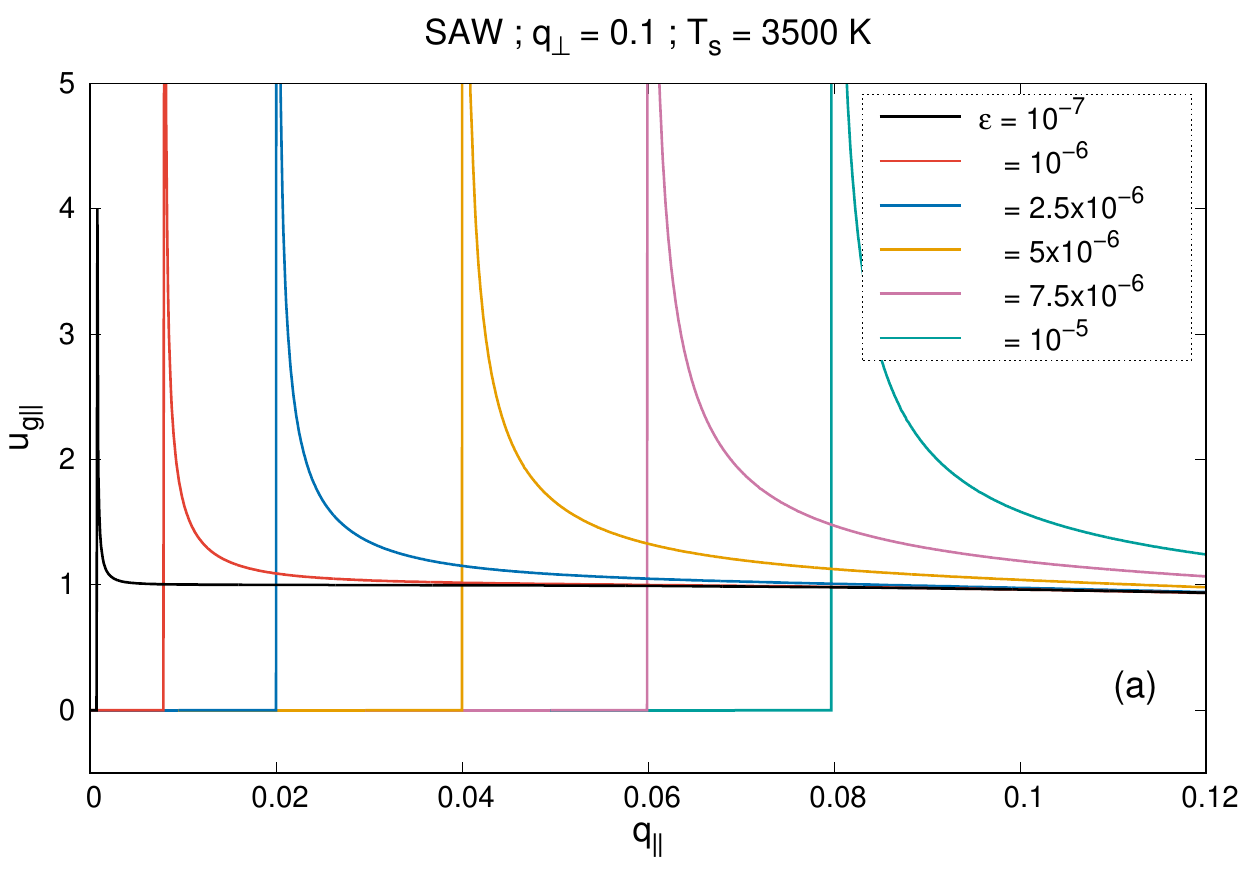}%
    \end{minipage}\vspace{\floatsep}
    \begin{minipage}[c]{\columnwidth}
        \centering
        \includegraphics[width=.99\columnwidth]{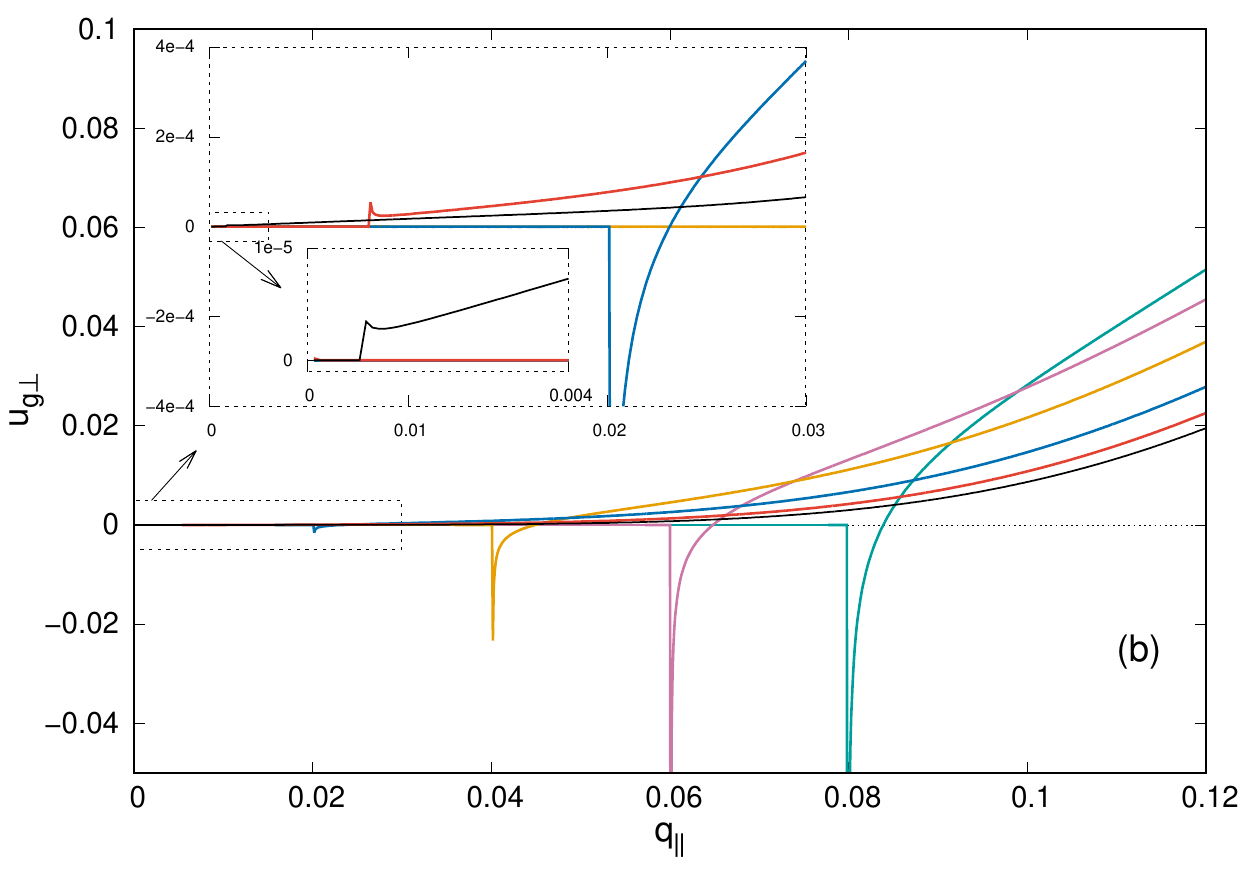}
    \end{minipage}
    \caption{(a) Parallel and (b) perpendicular component of normalized group velocity $u_g$ as functions of $q_\parallel$ for the SAWs presented in Fig.~\ref{fig:zr_q_Ts_eps}(a), with constant stellar surface temperature $T_s=3500\,$K and several values of dust to ion density ratio $\varepsilon$.}
    \label{fig:zr_q_eps_ioncyc}
\end{figure}

The group velocity values of the SAWs displayed in Fig.~\ref{fig:zr_q_Ts_eps}(a) are presented in Fig.~\ref{fig:zr_q_eps_ioncyc}. Both parallel (top panel) and perpendicular (bottom panel) components of $u_g$ show null values in the region where these modes have zero frequency $z_r$. At the point in which the modes become dispersive, we notice a jump of the group velocity from zero to a finite value, causing a discontinuity in the plot, and starts decreasing afterwards.

\begin{figure}
    \centering
    \includegraphics[width=\columnwidth]{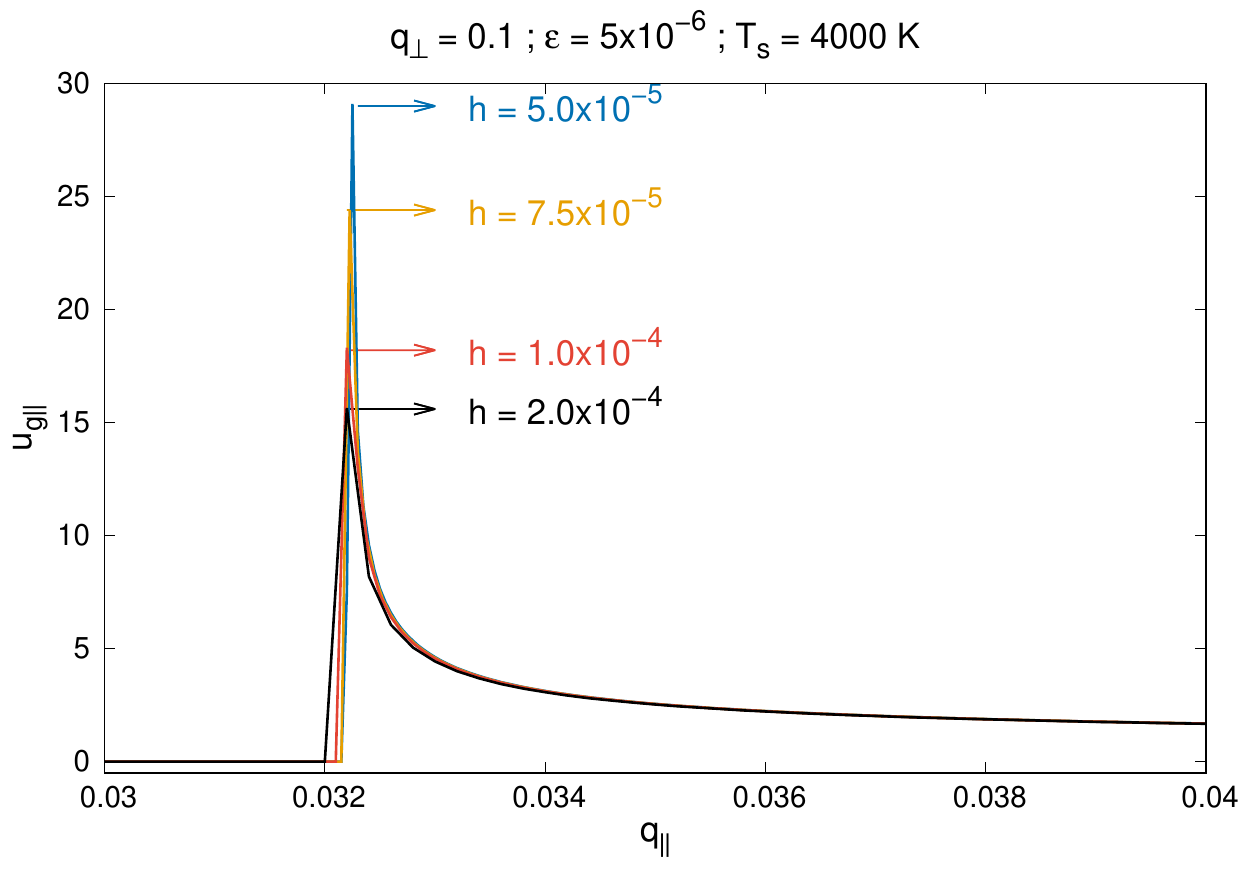}
    \caption{Parallel component of $u_g$ for SAW as a function of $q_\parallel$ for distinct values of the step $h$ utilised in the numerical differentiation (equation~\eqref{eq:numerical_derivative}).}
    \label{fig:zr_q_h}
\end{figure}

This value of the group velocity at the point of discontinuity is not well defined by our numerical method since the ion-cyclotron curves in Fig.~\ref{fig:zr_q_Ts_eps} indicate that the derivative at the point in which they no longer present $z_r=0$ tends to infinity (similarly to the behaviour of a square root function at the origin). For that reason, the value of $u_g$ at this point becomes dependent of the step $h$ utilised in equation \eqref{eq:numerical_derivative}, as is illustrated is Fig.~\ref{fig:zr_q_h}.

Nonetheless, all numerical derivatives presented in Fig.~\ref{fig:zr_q_h} show similar values of $u_{g\parallel}$ outside the vicinity of the discontinuity point. Anyhow, the fact that $u_g$ tends to infinity indicates that the group velocity in that spectral region does not have physical significance as the velocity of propagation of a wave packet, as it occurs in regions of anomalous dispersion \citep[see e.g.][]{brillouin_wave}.

Back to Fig.~\ref{fig:zr_q_eps_ioncyc}, we see that the parallel group velocity of SAWs starts decreasing after the point of discontinuity, converging to values near the Alfv{\'e}n velocity ($u_{g\parallel}=1$). The perpendicular component has values significantly lower and shows a distinct behaviour for some $\varepsilon$ values, with negative $u_{g\perp}$ right after the discontinuity point, i.e., in a small range of $q_\parallel$, the perpendicular group velocity for these waves is in opposite direction to the perpendicular phase velocity.

This region of negative $u_{g\perp}$ is rather small, turning into positive values for slightly bigger $q_\parallel$. For small dust density this feature does not appear, with the group velocity presenting positive values for all $q_\parallel$, as can be seen in the zoomed-in boxes in Fig.~\ref{fig:zr_q_eps_ioncyc}(b) for the $\varepsilon=10^{-6}$ and $10^{-7}$ curves.

\begin{figure}
    \centering
    \begin{minipage}[c]{\columnwidth}
        \centering
        \includegraphics[width=.99\columnwidth]{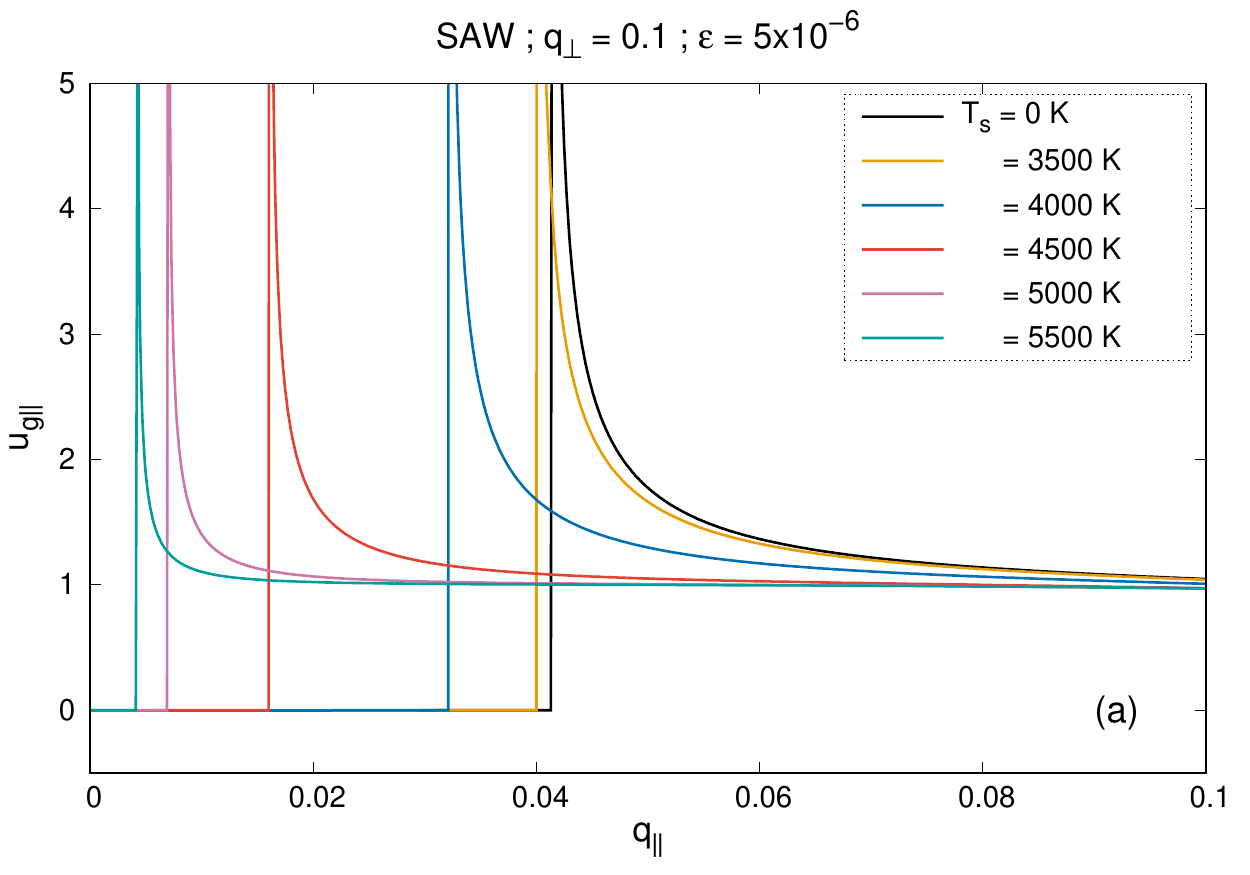}%
    \end{minipage}\vspace{\floatsep}
    \begin{minipage}[c]{\columnwidth}
        \centering
        \includegraphics[width=.99\columnwidth]{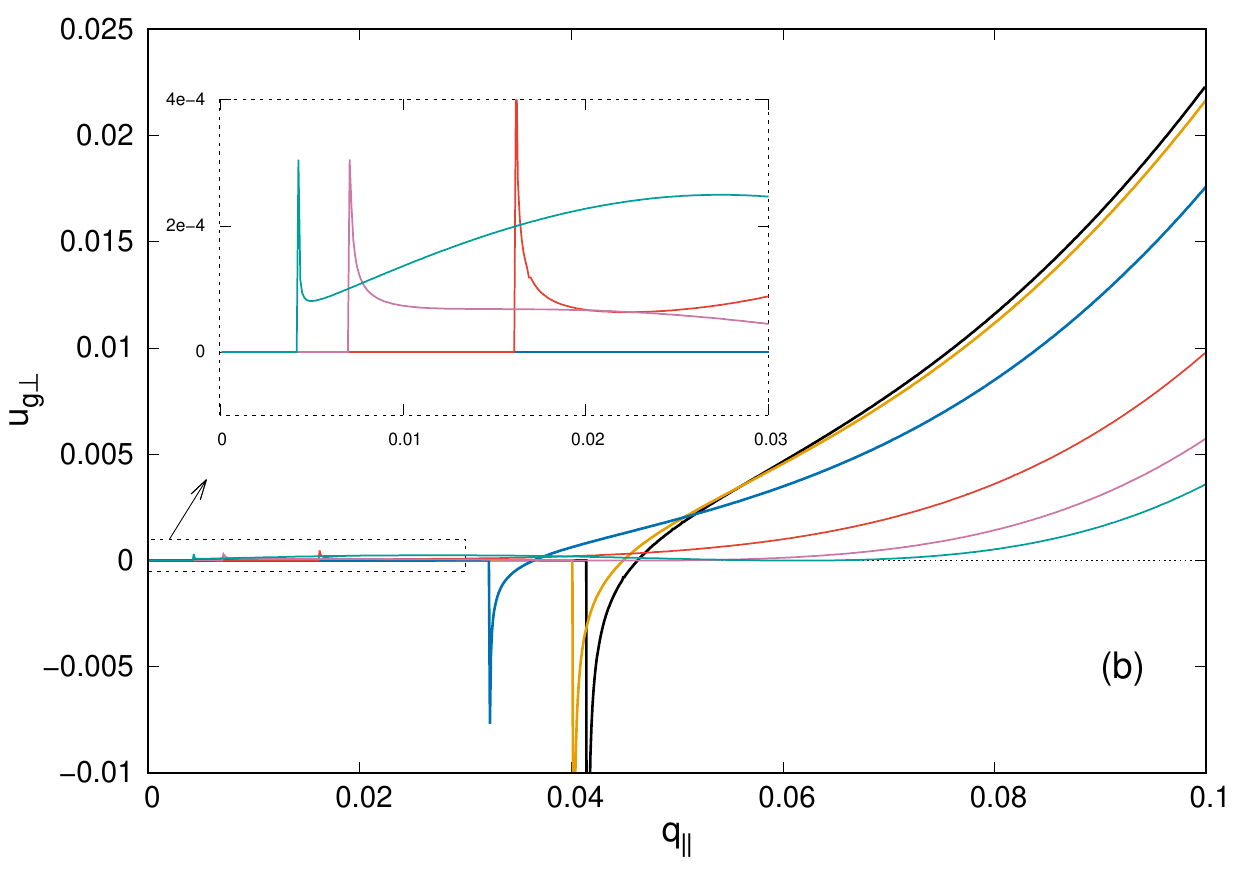}
    \end{minipage}
    \caption{(a) Parallel and (b) perpendicular component of normalized group velocity $u_g$ as functions of $q_\parallel$ for the SAWs presented in Fig.~\ref{fig:zr_q_Ts_eps}(b), with constant dust to ion density ratio $\varepsilon=5\times10^{-6}$ and several values of stellar surface temperature $T_s$.}
    \label{fig:zr_q_Ts_ioncyc}
\end{figure}

The situation in which dust density is maintained fixed and the radiation flux is changed through the stellar surface temperature $T_s$ is shown in Fig.~\ref{fig:zr_q_Ts_ioncyc} for SAW and same set of parameters as in Fig.~\ref{fig:zr_q_Ts_eps}(b). The behaviour of the parallel component of group velocity is very similar to what is observed in Fig.~\ref{fig:zr_q_eps_ioncyc}(a), with $u_{g\parallel}$ showing a region of anomalous dispersion at the point where the modes start presenting non-zero values of $z_r$, and decreasing to a value near the Alfv{\'e}n velocity outside the vicinity of the discontinuity point.

The perpendicular component $u_{g\perp}$ of these modes present, as before, much smaller values than the parallel component and situations with negative values in a short interval of $q_\parallel$ near the point of discontinuity, for small radiation flux. For greater values of $T_s$, the perpendicular group velocity is positive (see zoomed-in box in bottom panel) right after the discontinuity point, as occurs with the parallel component.

\begin{figure}
    \centering
    \includegraphics[width=\columnwidth]{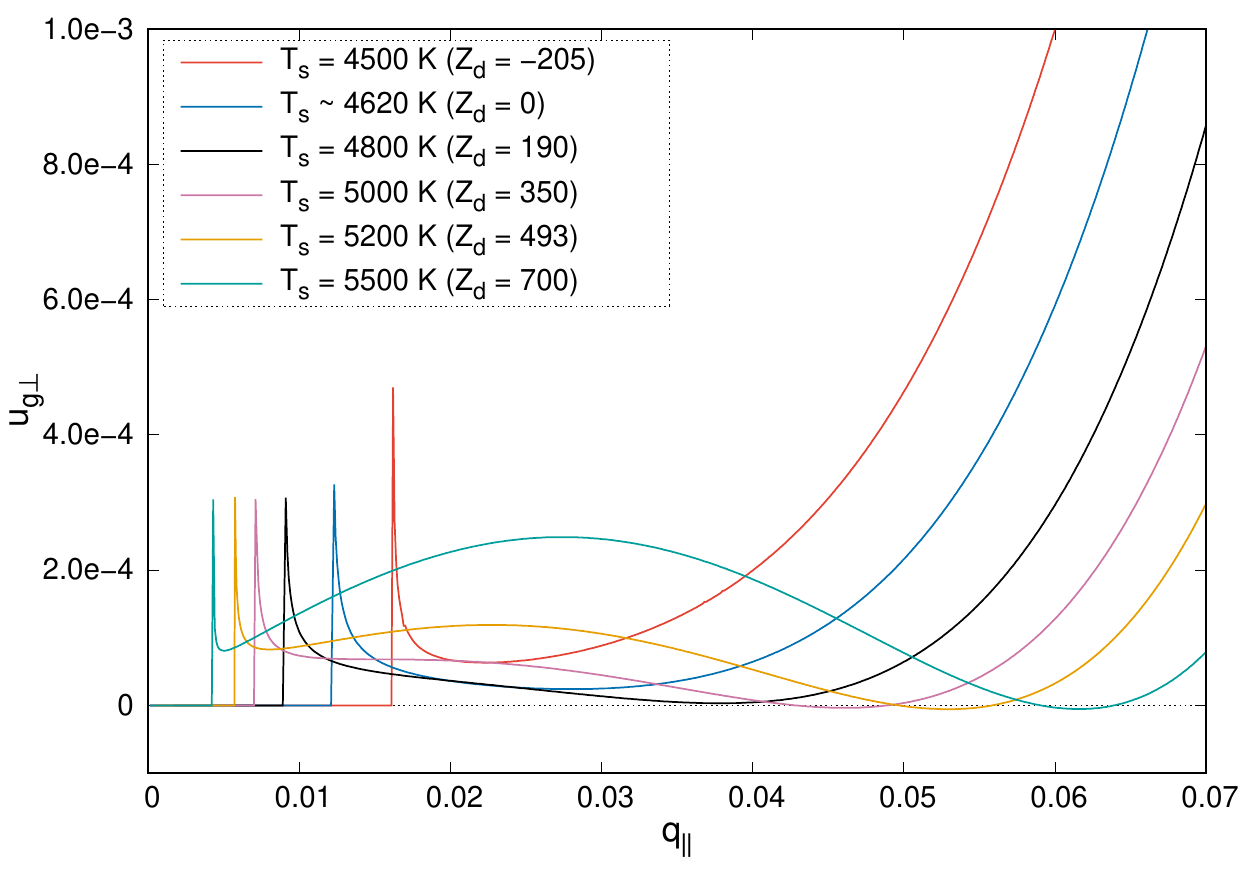}
    \caption{Perpendicular component of normalized group velocity $u_g$ as function of $q_\parallel$ for the SAWs, with constant dust to ion density ratio $\varepsilon=5\times10^{-6}$, $q_\perp=0.1$, and several values of stellar surface temperature $T_s$, along with their corresponding dust charge number $Z_\mathrm{d}$.}
    \label{fig:zr_q_Ts_ioncyc_velperp}
\end{figure}

However, we notice in some situations that $u_{g\perp}$ presents a complex behaviour and may even decrease to negative values in a short interval of $q_\parallel$, and become positive again. This can be visualised in Fig.~\ref{fig:zr_q_Ts_ioncyc_velperp}, in which several other values of $T_s$ have been added, in order to better show the transition of a curve with positive values for all $q_\parallel$ ($T_s \lesssim 4800$\,K) to the cases showing negative $u_{g\perp}$ in some region ($T_s \gtrsim 5000$\,K).

\section{Conclusions} \label{sec:conclusions}

The characteristics of the group velocity of Alfv{\'e}n waves propagating obliquely to the direction of the ambient magnetic field in a homogeneous dusty plasma are investigated. For that, we used a kinetic formulation which considers Maxwellian distributions for ions and electrons, and immobile dust particles. The dust grains are considered to have same radius and are charged by the absorption of plasma particles via inelastic collisions, and by photoionization due to the radiation flux coming from the star's surface. The parameters considered are typical of a stellar wind from a carbon-rich star.

The study is performed for the normal wave modes that arise from the dispersion relation, namely the CAW (or whistler, for large wavenumber) and SAW (or ion-cyclotron). We observe the changes in the modes' normalized group velocity $u_g$ along and transverse to the magnetic field by varying parameters related to the dust population, such as dust to ion number density ratio and equilibrium electrical charge of the dust grains (by varying the radiation flux via the stellar surface temperature).

The results show that, for a fixed transverse component of the normalized wavenumber $q$, the parallel group velocity of CAWs generally increases with $q_\parallel$, whilst its perpendicular component decreases, i.e., the direction of the wave packet propagation is greatly modified with the wave-vector direction. We also noticed that, for small $q_\parallel$, the increase of dust number density can greatly increase the $u_{g\perp}$ values whilst maintaining $u_{g\parallel}$ almost constant, indicating that in dusty plasmas these waves with large wavelength will propagate energy at greater angles relative to the magnetic field, in comparison with conventional plasmas.

SAWs, on the other hand, will generally present much higher values of $u_{g\parallel}$, consequence of its anisotropic nature, with group velocity propagating practically along the magnetic field. Nevertheless, when these waves have non zero frequency, its perpendicular component shows to be non-vanishing and, for sufficiently high values of dust number density, it may present negative values in a small interval of $q_\parallel$. This interval with negative $u_{g\perp}$ may be modified, or even vanish, by changes in the radiation flux incident on the dust particles, which modifies the grains' equilibrium electrical charge.

From the present study and from the previous publication \citep[][]{detoni2022}, it is seen that obliquely propagating Alfv{\'e}n waves are significantly changing their properties in a magnetized dusty plasma with the modification of dust related parameters. These findings may prove to be helpful in better understanding plasma heating and energy flux processes in stellar wind environments, since the plasma parameters in these places could be greatly modified by the presence of dust. For instance, waves generated within different distances from the star's radiating surface will have distinct characteristics, given that dust number density and radiation flux are dependent of the distance from the star.

Further studies can be followed from this work. Our model incorporates only one population of dust grains with constant size. However, it is expected that space dusty plasmas present several dust populations with different sizes, with radii varying from about $10^{-4}$ to $10^{-7}$\,cm in a dust-driven wind \citep[][]{kruger1997two}. The work of \citet{galvao_ziebell2012} included, within the kinetic theory, a discrete distribution of grain sizes to derive an expression of the dielectric tensor for a magnetized dusty plasma. In the near future, we intend to make use of that formalism to incorporate a continuous distribution of grain sizes to our model, given that in space environments the dust radii are often described by a continuous distribution function \citep[see e.g.][]{mathis1977size,Dominik_1989,Hoefner_1992}. 

Moreover, our model does not take into account non-linear effects of the system. The development of a quasi-linear theory for a dusty plasma would allow to include the dynamical evolution of the grains' electrical charge in a self-consistent way in the kinetic theory. Also, a better understanding of the wave-particle and wave-wave interactions could be accomplished by a non-linear treatment, such as the weak turbulence theory of plasmas \citep[][]{galvao2015}. These non-linear kinetic processes are believed to be important in the mechanism responsible for the heating and acceleration of stellar winds \citep[][]{dePontieu2007chromospheric,Cranmer+15/04,Raouafi+21/04}. We also intend to pursue this line of investigation.

\section*{Acknowledgements}

This study was financed in part by the Coordena{\c{c}}{\~a}o de Aperfei{\c{c}}oamento de Pessoal de N{\'i}vel Superior – Brasil (CAPES) – Finance Code 001. RG acknowledges support from CNPq (Brazil), grant No. 307845/2018-4. LFZ acknowledges support from CNPq (Brazil), grant No. 302708/2018-9.

\section*{Data Availability}

The data underlying this article will be shared on reasonable request to the corresponding author.



\bibliographystyle{mnras}
\bibliography{MAIN} 








\bsp	
\label{lastpage}
\end{document}